\documentclass {aastex631}

    {\pagebreak[4]\global\pdfpageattr\expandafter{\the\pdfpageattr/Rotate 90}}%
    {\pagebreak[4]\global\pdfpageattr\expandafter{\the\pdfpageattr/Rotate 0}}%

\usepackage{natbib, graphicx, morefloats, multirow, pbox, booktabs, makecell}
\usepackage{hyperref}
\usepackage{soul}

\begin{document}

\title{Observations of DNC and DCO$^+$ toward the $\int$-shaped Filament and Starless Cores in the Orion Molecular Clouds}

\correspondingauthor{Ken'ichi Tatematsu}
\email{kenichi.tatematsu@nifty.com}

\author[0000-0002-8149-8546]{Ken'ichi Tatematsu}
\affil{Nobeyama Radio Observatory, National Astronomical Observatory of Japan,
National Institutes of Natural Sciences,
Nobeyama, Minamimaki, Minamisaku, Nagano 384-1305, Japan}
\affiliation{Astronomical Science Program,
SOKENDAI (The Graduate University for Advanced Studies),
2-21-1 Osawa, Mitaka, Tokyo 181-8588, Japan}
\affiliation{kenichi.tatematsu@nifty.com}

\author[0000-0003-0732-2937]{Atsushi Nishimura}
\affiliation{Nobeyama Radio Observatory, National Astronomical Observatory of Japan,
National Institutes of Natural Sciences,
Nobeyama, Minamimaki, Minamisaku, Nagano 384-1305, Japan}
\affiliation{Astronomical Science Program,
SOKENDAI (The Graduate University for Advanced Studies),
2-21-1 Osawa, Mitaka, Tokyo 181-8588, Japan}

\author[0000-0002-8686-6862]{Hideo Ogawa}
\affiliation{Department of Physics, Graduate School of Science, Osaka Metropolitan University, 3-3-138 Sugimoto, Sumiyoshi-ku, Osaka 558-8585, Japan}

\author[0000-0002-3297-4497]{Nami Sakai}
\affiliation{Star and Planet Formation Laboratory, RIKEN, 2-1 Hirosawa, Wako, Saitama, 351-0198, Japan}

\author[0000-0003-4521-7492]{Takeshi Sakai}
\affiliation{Graduate School of Informatics and Engineering, The University of Electro-Communications, Chofu, Tokyo 182-8585, Japan}

\author[0000-0002-2062-1600]{Kazuki Tokuda}
\affiliation{Faculty of Education, Kagawa University, Saiwai-cho 1-1, Takamatsu, Kagawa 760-8522, Japan}
\affiliation{Department of Earth and Planetary Sciences, Faculty of Science, Kyushu University, Nishi-ku, Fukuoka 819-0395, Japan}
\affiliation{National Astronomical Observatory of Japan, National Institutes of Natural Sciences, 2-21-1 Osawa, Mitaka, Tokyo 181-8588, Japan}

\author[0000-0002-0164-3572]{Yutaka Hasegawa}
\affiliation{National Institute of Information and Communications Technology,
4-2-1, Nukuikitamachi, Koganei-shi, Tokyo 184-8795, Japan}
\affiliation{Department of Physics, Graduate School of Science, Osaka Metropolitan University, 3-3-138 Sugimoto, Sumiyoshi-ku, Osaka 558-8585, Japan}

\author[0000-0001-5859-8840]{Yasumasa Yamasaki}
\affil{National Astronomical Observatory of Japan,
National Institutes of Natural Sciences,
2-21-1 Osawa, Mitaka, Tokyo 181-8588, Japan}
\affiliation{Department of Physics, Graduate School of Science, Osaka Metropolitan University, 3-3-138 Sugimoto, Sumiyoshi-ku, Osaka 558-8585, Japan}

\author[0000-0001-7826-3837]{Toshikazu Onishi}
\affiliation{Department of Physics, Graduate School of Science, Osaka Metropolitan University, 3-3-138 Sugimoto, Sumiyoshi-ku, Osaka 558-8585, Japan}

\author[0000-0001-9304-7884]{Naomi Hirano}
\affiliation{Institute of Astronomy and Astrophysics, Academia Sinica, No.1, Sec. 4, Roosevelt Rd, Taipei 106319, Taiwan (R.O.C.)}

\author[0000-0003-4603-7119]{Sheng-Yuan Liu}
\affiliation{Institute of Astronomy and Astrophysics, Academia Sinica, No.1, Sec. 4, Roosevelt Rd, Taipei 106319, Taiwan (R.O.C.)}

\author[0000-0002-5286-2564]{Tie Liu}
\affiliation{Shanghai Astronomical Observatory, Chinese Academy of Sciences, 80 Nandan Road, Shanghai 200030, P. R. China}
\affiliation{Korea Astronomy and Space Science Institute,
776 Daedeok-daero, Yuseong-gu, Daejeon 34055, South Korea}
\affiliation{East Asian Observatory, 660 N. A'ohoku Place, Hilo, HI 96720, USA}

\author[0000-0002-2338-4583]{Somnath Dutta}
\affiliation{Institute of Astronomy and Astrophysics, Academia Sinica, No.1, Sec. 4, Roosevelt Rd, Taipei 106319, Taiwan (R.O.C.)}

\author[0000-0002-4393-3463]{Dipen Sahu}
\affiliation{Physical Research laboratory, Navrangpura, Ahmedabad, Gujarat 380009, India}
\affiliation{Institute of Astronomy and Astrophysics, Academia Sinica, No.1, Sec. 4, Roosevelt Rd, Taipei 106319, Taiwan (R.O.C.)}

\author[0000-0002-3024-5864]{Chin-Fei Lee}
\affiliation{Institute of Astronomy and Astrophysics, Academia Sinica, No.1, Sec. 4, Roosevelt Rd, Taipei 106319, Taiwan (R.O.C.)}

\author[0000-0003-2412-7092]{Kee-Tae Kim}
\affiliation{Korea Astronomy and Space Science Institute,
776 Daedeok-daero, Yuseong-gu, Daejeon 34055, Republic of Korea}
\affiliation{University of Science and Technology, Korea (UST), 217 Gajeong-ro, Yuseong-gu,
Daejeon 34113, Republic of Korea}

\author[0000-0001-8509-1818]{Gary A. Fuller}
\affiliation{Jodrell Bank Centre for Astrophysics, School of Physics and Astronomy, University of Manchester, Oxford Road, Manchester, M13 9PL, UK}

\author[0000-0002-1369-1563]{Shih-Ying Hsu}
\affiliation{Institute of Astronomy and Astrophysics, Academia Sinica, No.1, Sec. 4, Roosevelt Rd, Taipei 106319, Taiwan (R.O.C.)}

\author[0000-0003-0537-5461]{Hee-Weon Yi}
\affiliation{Korea Astronomy and Space Science Institute,
776 Daedeok-daero, Yuseong-gu, Daejeon 34055, Republic of Korea}

\author[0000-0002-6264-6507]{Sho Masui}
\affil{National Astronomical Observatory of Japan,
National Institutes of Natural Sciences,
2-21-1 Osawa, Mitaka, Tokyo 181-8588, Japan}
\affiliation{Department of Physics, Graduate School of Science, Osaka Metropolitan University, 3-3-138 Sugimoto, Sumiyoshi-ku, Osaka 558-8585, Japan}

\author{Shimpei Nishimoto}
\affiliation{Department of Physics, Graduate School of Science, Osaka Metropolitan University, 3-3-138 Sugimoto, Sumiyoshi-ku, Osaka 558-8585, Japan}

\author{Chieko Miyazawa}
\affil{Nobeyama Radio Observatory, National Astronomical Observatory of Japan,
National Institutes of Natural Sciences,
Nobeyama, Minamimaki, Minamisaku, Nagano 384-1305, Japan}

\author{Toshikazu Takahashi}
\affil{Nobeyama Radio Observatory, National Astronomical Observatory of Japan,
National Institutes of Natural Sciences,
Nobeyama, Minamimaki, Minamisaku, Nagano 384-1305, Japan}

\author{Jun Maekawa}
\affil{Nobeyama Radio Observatory, National Astronomical Observatory of Japan,
National Institutes of Natural Sciences,
Nobeyama, Minamimaki, Minamisaku, Nagano 384-1305, Japan}

\author[0000-0002-7468-659X]{Alvaro Gonzalez}
\affil{Joint ALMA Observatory, Alonso de C\'{o}rdova 3107, Vitacura , Santiago, Chile}
\affil{National Astronomical Observatory of Japan,
National Institutes of Natural Sciences,
2-21-1 Osawa, Mitaka, Tokyo 181-8588, Japan}

\author[0000-0001-7949-6528]{Takafumi Kojima}
\affil{National Astronomical Observatory of Japan,
National Institutes of Natural Sciences,
2-21-1 Osawa, Mitaka, Tokyo 181-8588, Japan}

\author[0000-0003-0822-2591]{Keiko Kaneko}
\affil{National Astronomical Observatory of Japan,
National Institutes of Natural Sciences,
2-21-1 Osawa, Mitaka, Tokyo 181-8588, Japan}

\author{Ken Miyato}
\affiliation{Graduate School of Informatics and Engineering, The University of Electro-Communications, Chofu, Tokyo 182-8585, Japan}

%\author{TRAO Key Science Program ``TOP'' collaboration}

\begin{abstract}
Although the deuterium fraction is known to be a powerful evolutionary tracer, 
its variation within individual molecular cloud cores is still poorly understood.
The northern $\int$-shaped filament and 20 individual starless cores in the Orion A and B clouds
were mapped in the deuterated molecules of DNC and DCO$^+$ with the Receiver 7BEE installed on
the Nobeyama 45~m radio telescope.
In a $\sim 5\arcmin\times30\arcmin$ map of the northern $\int$-shaped filament
in the Orion A cloud,
the DNC emission is detected over the filament, whereas the DCO$^+$ emission is
localized toward OMC-3, the northernmost region of the filament.
The difference in distribution between DNC and DCO$^+$
can be attributed to that between N- and C-bearing molecules
as previously suggested by Tatematsu et al.
High DNC/HN$^{13}$C column density ratios were observed in OMC-2 and OMC-3, and low ratios in OMC-1.
It seems that OMC-2 and OMC-3 still contain molecular gas close to
the onset of star formation.
In 3${\arcmin}~\times~3{\arcmin}$ maps of the individual starless cores in Orion, 
the column density ratios of DNC/HN$^{13}$C
and DCO$^+$/H$^{13}$CO$^+$ are found to be rather constant locally 
within each core, although the core-to-core variation is not small.
Similar timescales of deuterization, depletion, and dynamical evolution
might explain the locally constant ratio.
\end{abstract}


\section{Introduction} \label{sec:intro}
%\section{INTRODUCTION}

The evolutionary stage of a molecular cloud core prior to star formation (a starless core) is essential for understanding 
the physical processes leading to star formation, but it is harder to assess than that of a core that has already formed a protostar 
(a star-forming core or a protostellar core). 
For example, the evolutionary stage of a protostar can be characterized by its bolometric temperature \citep{1995ApJ...445..377C}. 
A prestellar core is a type of starless core thought to be close to the onset of 
star formation and characterized by a steep radial density distribution, 
indicating that self-gravity dominates.
Although the evolutionary stage estimate methods are rather limited
\citep{1992ApJ...392..551S,2014PASJ...66...16T,2020ApJ...895..119T}, 
deuterium fractionation in molecules such as N$_2$H$^+$, HNC, and HCO$^+$,
which are formed in the gas phase,
provides a powerful chemical clock for the early phases of star formation, 
because it sensitively traces the combined effects of temperature, 
density, CO depletion, and ion–molecule chemistry.
The evolutionary behavior of deuterium fractionation can depend on molecular species because of different chemical pathways 
\citep{2012A&ARv..20...56C,2014prpl.conf..859C,2015A&A...575A..87F}. 
Nevertheless, the deuterium fraction in N$_2$H$^+$ and HNC tends to increase during the prestellar phase 
and reaches high values immediately prior to the onset of star formation \citep{2002ApJ...565..344C,2005ApJ...619..379C,2006ApJ...646..258H,2009A&A...496..731E}. 
It then decreases after the onset of star formation in both low-mass and high-mass star-forming regions
\citep{2009A&A...496..731E,2012ApJ...747..140S,2014MNRAS.440..448F,2018ApJ...869...51I}.
Characteristics of the deuterium fraction in massive star-forming regions can be similar to 
those in low-mass star-forming regions for some molecular tracers \citep{2011A&A...529L...7F}. 
However, significant differences have also been reported, likely reflecting different physical conditions 
and chemical pathways in high-mass and intermediate-mass environments \citep[e.g.,][]{2015ApJ...814...31K,2020MNRAS.496.1990R}.
The Orion molecular clouds (OMC) are a prototypical massive star-forming complex, but they also host rich populations of low- and intermediate-mass protostars, 
such as in the OMC-2 and OMC-3 regions.
Therefore, a wide-field, uniform mapping of deuterium fractionation across Orion is essential to connect chemical evolution 
with the global evolutionary sequence of the clouds. In this paper, we focus on the low- and intermediate-mass star-forming regions within Orion, where deuterium fractionation is expected to be a sensitive evolutionary tracer.
DNC and DCO$^+$ are particularly powerful tracers because they probe complementary aspects of chemical evolution: DNC traces nitrogen chemistry 
and cold, chemically evolved gas, 
whereas DCO$^+$ is closely linked to CO depletion and ion–molecule chemistry, 
making them ideal probes of the initial conditions of star formation.

In low-mass or intermediate-mass star-forming regions, single-dish DNC observations toward regions including starless clumps/cores have been made by, for example,
\citet{2001ApJ...547..814H,2003ApJ...594..859H}, who mapped the
DNC emission over a $5\arcmin\times10\arcmin$ toward TMC-1 etc. 
on a $1\arcmin$ grid by using the Nobeyama 45~m telescope. 
\citet{2019PASJ...71S..10N}
obtained a DNC map toward Orion Molecular Cloud 2 FIR 4 ($3\arcmin\times3\arcmin$ area).
\citet{2018ApJ...869...51I} conducted a deuterium fraction survey in DNC/HN$^{13}$C toward
protostellar sources in the Perseus molecular cloud,
and found a negative correlation between deuterium fraction and the bolometric temperature.
\citet{2023A&A...676A..78G} found similar deuterium fractions in
DNC/HN$^{13}$C between the starless core L1544 and the protostellar core HH211.
Regarding DCO$^+$, single-dish observations toward regions containing starless clumps/cores have been carried
out  by \citet{1982ApJ...255..160W}, \citet{1990ApJ...365..269L}, and \citet{1995ApJ...448..207B}.
\citet{2002ApJ...565..344C}
studied
DCO$^+$ in L1544 ($2\farcm2\times2\farcm2$ area)
using the IRAM 30~m telescope, and derived the ionization degree.
\citet{2024ApJ...963...12T}
conducted
mapping observations of DCN and DCO$^+$ toward Orion KL
($1\farcm5\times1\farcm5$ area) and argued that the 
DCN/DCO$^+$ column density ratio 
depends on density.  However, previous studies have been limited to single-point observations or small fields, and no uniform mapping study 
has yet investigated the large-scale distribution of DNC and DCO$^+$ 
across the prototypical giant molecular cloud complex, OMC, that also hosts low- and intermediate-mass star formation.

The physical and chemical properties of starless cores provide important constraints on the initial conditions for star formation, particularly on the transition from stable to unstable cores \citep[e.g.,][]{2022ApJ...931...33T}.
\citet{2023ApJ...945..156S} investigated the density structure of centrally concentrated prestellar cores and showed that they exhibit steep radial density profiles consistent with dominant self-gravity.
\citet{2025ApJ...984L..58H} reported the detection of turbulence-induced mass-assembly shocks in starless cores.
On the theoretical side, \citet{2003ApJ...593..906A} modeled the radial distributions of various molecules, including deuterated species, for different collapse scenarios.

We employed the seven-element receiver 7BEE
\citep[7-BEam Equipment for the Nobeyama 45~m Telescope,][]{2023PASJ...75..499Y} 
on the Nobeyama 45 m telescope, which enables sensitive, wide-field, fully sampled mapping of deuterated molecules in the 70–80~GHz band.
Because only a limited number of facilities operating in this frequency range, 7BEE provides a rare opportunity to investigate deuterium fractionation.
Since we aim to study variations of the deuterium fraction within individual molecular cloud cores, high mapping capability is required because the emission from deuterated molecules is generally weak.
To date, observations of deuterated molecules have been limited by the map size 
($<10\arcmin$ square), the number of maps, and the sampling (some large maps were undersampled).
In this study, 7BEE was used for observations of deuterated molecules in the 70–80~GHz band, while the receiver FOREST was used for observations of the corresponding non-deuterated molecules in the 80–90~GHz band.
The distance to the Orion regions is assumed to be
420 pc
\citep{2017ApJ...834..142K,2019MNRAS.487.2977G}.

\section{Observations} \label{sec:obs}

We carried out mapping observations with the Nobeyama
\footnote{Nobeyama Radio Observatory
is a branch of the National Astronomical Observatory of Japan,
National Institutes of Natural Sciences} 
45 m radio telescope toward the Orion A and B clouds.
The Orion A cloud, a $\sim$40~pc-long giant molecular cloud hosts
the major star-forming sites, OMC-1, OMC-2, and OMC-3
\citep[e.g.,][]{1974ApJ...191L.121G,1976ApJ...209..452K,1993A&A...270..468D,1997ApJ...474L.135C}
aligned along the $\sim$10~pc-long $\int$-shaped filament.
Our observations cover the northern $\sim$4~pc of the $\int$-shaped filament 
($\sim5\arcmin\times30\arcmin$ area), 
together with 20 starless SCUBA-2 cores distributed in the Orion A and B clouds.
The observed starless SCUBA-2 cores (cores 04, 10$-$12, 14, 15, 17$–$25, 27–30, and 32) 
were selected from \citet{2022ApJ...931...33T}, which was conducted as part of the TOP-SCOPE collaboration
\citep{2015PKAS...30...79L,2018ApJS..234...28L,2017ApJS..228...12T,2021ApJS..256...25T,2018ApJS..236...51Y,2019MNRAS.485.2895E,2024MNRAS.530.5192E,2020ApJS..249...33K}
and
the ALMASOP collaboration 
\citep{2020ApJS..251...20D,2021ApJ...907L..15S,2023ApJ...945..156S,2025ApJ...984L..58H,2026ApJ...997...16H}.
The TOP-SCOPE collaboration identified dense cores in the 850~$\mu$m continuum with SCUBA-2 
toward the \textit{Planck} Galactic Cold Clumps (PGCCs) from the \textit{Planck} all-sky survey 
\citep{2011AA...536A..23P,2016AA...594A..28P}. 
Out of the 39 starless cores in Orion in our sample, we observed the same 20 cores that were included in our previous survey of inward motions.
The JCMT SCUBA-2 core names and their coordinates 
are adopted from \citet{2018ApJS..236...51Y}. 
Cores lacking young stellar objects (YSOs) are termed starless cores.
The classification of the starless cores follows \citet{2020ApJS..249...33K}, 
who conducted observations with the same radio telescope at the same spatial resolution.

For the receiver frontend for the deuterated molecules, 7BEE was employed.
The receiver has seven element beams, and
angular distances between the central beam and the surrounding six beams were
$70\arcsec$.
The half-power beam width (HPBW) and main-beam efficiency $\eta_{mb}$ at 74 GHz with 7BEE were 
$21\farcs4~\pm~0\farcs7$ (0.044~pc at 420~pc distance) and $42.3~\pm~4.8\%$, respectively. 
In front of the feed horns, the Kapton film window shield was used as vacuum windows.
7BEE was designed to observe simultaneously two out of the pre-selected candidate molecular lines of
the $J = 1-0$ transitions of DNC, DCO$^+$, DCN, N$_2$D$^+$, H$^{13}$CN, H$^{13}$CO$^+$, HN$^{13}$C, HCN, HCO$^+$, HNC, N$_2$H$^+$, C$^{18}$O, $^{13}$CO, 
and $^{12}$CO, the $v = 1, J = 2-1$ transition of SiO (for commissioning), etc.
For the present study, the $J = 1-0$ transitions of DNC and DCO$^+$
\citep[76.3057270 and 72.0393028 GHz, respectively;][]{1998JQSRT..60..883P}
were simultaneously observed with 7BEE.

For the non-deuterated molecules, the Receiver FOREST \citep[FOur-beam REceiver System on the 45~m Telescope;][]{2016SPIE.9914E..1ZM} was used.
The HPBW and $\eta_{mb}$ at 86~GHz were $18\farcs4~\pm~0\farcs2$ (0.037~pc at a distance of 420~pc) and 
$50.6~\pm~3.6\%$, respectively.
The $J=1-0$ transitions of HN$^{13}$C, HNC, H$^{13}$CO$^+$, and HCO$^+$ were observed simultaneously.
We newly mapped the northern $\int$-shaped filament, while the data toward the starless cores were taken from \citet{2022ApJ...931...33T} to avoid duplicate observations.
The same region as that observed with 7BEE was mapped with FOREST.
 
For the receiver backend for both the receivers, 
the SAM45 \citep[Spectral Analysis Machine for the 45~m telescope, ][]{2012PASJ...64...29K} 
was employed with a channel separation of 30.52~kHz, which corresponded to $\sim$0.1~km~s$^{-1}$.

Observations with 7BEE were carried out in 2023 Mar, Apr, Nov, and Dec.
Those with FOREST were done in 2023 Feb and Dec.
We noticed that refrigerated HEMT amplifiers employed in 7BEE 
had been gradually degraded
from the initial commissioning (2022 Sep-2023 Mar) after the installation on the telescope (2022 Aug 2).
In 2023 Mar and Apr, 22 arrays out of the 28 arrays 
(7 beams $\times$ 2 polarizations $\times$ 2 lines) were available for observations
(four arrays were suffered from the degraded HEMT amplifiers 
and 2 arrays had wrong IF cabling).
In 2023 Nov and Dec, 13 arrays out of the 28 arrays were 
available for observations
(seven arrays were suffered from the degraded HEMT amplifiers and eight arrays were affected by malfunction in the SAM45 spectrometers).
The degradation of the HEMT amplifiers was suspected to be caused by humidity inside the dewar originating from the Kapton vacuum windows.
On-the-fly (OTF) mapping observations \citep{2008PASJ...60..445S} 
were mostly performed in both the R.A. and decl. directions
toward the $\int$-shaped filament and the individual starless cores,
to minimize striping effects, but in some cases, 
scans of only one direction were made.
The map center for the individual starless cores is taken to be the core center.
The map size for the individual cores with 7BEE defined with the scan pattern of the central beam was 5${\arcmin}~\times~5{\arcmin}$,
but the inner 3${\arcmin}~\times~3{\arcmin}$ map with almost constant sensitivity 
is used in the current study; 
outside the inner region, the number of the employed beam is smaller, 
and the sensitivity is worse.
The HN$^{13}$C and H$^{13}$CO$^+$ map with FOREST 
has the same map size.
Our observations fully cover and spatially resolve the molecular cloud cores.
The position switching mode was employed, and
the reference off positions were ($\Delta$R.A., $\Delta$decl.) = 
$(-30\arcmin, 5\arcmin), (-30\arcmin, 0\arcmin), (20\arcmin, 20\arcmin), (-20\arcmin, -20\arcmin)$,
and $(20\arcmin, 20\arcmin)$ from Orion KL and the map centers, for
the $\int$-shaped filament, cores 4$-$12, 14$-$15, 17$-$18, and 19$-$32, respectively,
which were confirmed to be free of line emission.
The map reference center for the $\int$-shaped filament was
(R.A., decl.) (J2000.0) = ($5^h 35^m 14\fs5, -5\degr 22\arcmin 30\arcsec$).
Those for the individual starless cores except cores 12, 21, and 25
are the SCUBA-2 continuum peaks.
Cores 12, 21, and 25  are adjacent to cores 11, 20, and 24, respectively,
and were mapped simultaneously.
The typical system temperature was approximately 
200 K for both 7BEE and FOREST.
The sensitivity obtained with FOREST is comparable to that obtained with 7BEE for the $\int$-shaped filament, and better for the individual starless cores. This difference is mainly due to differences in integration time.
The telescope pointing calibration was established at 1.0$-$1.5 h intervals toward the SiO maser source, Orion KL, which resulted in a pointing accuracy of $\lesssim5\arcsec$. 
Linear baselines were subtracted from the spectral data, 
and the data were stacked on a grid of 
6 $\arcsec$ with the Bessel$-$Gauss function in the NOSTAR program
of the Nobeyama Radio Observatory \citep{2008PASJ...60..445S}. 
The line intensity of all the observations was expressed in terms of the main-beam radiation temperature $T_{mb}$,
which is 
the antenna temperature $T_{\rm A}^{\ast}$ corrected for atmospheric extinction using standard chopper wheel calibration, and finally divided by the main-beam efficiency.

\section{RESULTS \label{sec:res}}

\subsection{Distribution}

\begin{figure}
\includegraphics[bb=0 0 350 450, width=15cm]{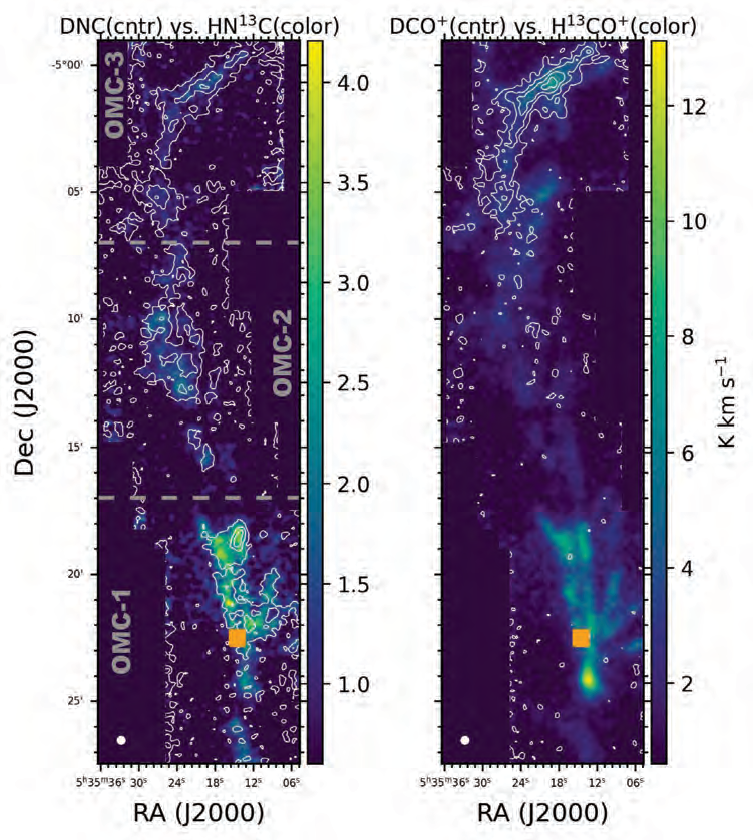}
\caption{
Contours represent the DNC (left) and DCO$^+$ (right) distributions toward the northern half of the $\int$-shaped filament
in the Orion A cloud.
The velocity integration ranges were chosen to encompass the detected emission: $V_{LSR}$ = 9.7–11.9 km s$^{-1}$ 
and 7.1–12.5 km s$^{-1}$ for DNC and DCO$^+$, respectively.
The color shows the HN$^{13}$C (left) and H$^{13}$CO$^+$ (right) distributions for comparison with the corresponding deuterated molecules.
The velocity integration ranges for HN$^{13}$C and H$^{13}$CO$^+$ are $V_{LSR}$ = 4.4–13.2 km s$^{-1}$ and 5.3–12.3 km s$^{-1}$, respectively.
The color scale starts at 0.6 K km s$^{-1}$ (3$\sigma$), and the contour interval is 1 K km s$^{-1}$ (5$\sigma$).
The horizontal gray dashed lines mark the boundaries between OMC-1, OMC-2, and OMC-3.
The orange filled box indicates the location of Orion KL.
The white filled circle in the lower-left corner represents the HPBW for the deuterated molecules.
\label{fig:OADNC40+DCO40}}
\end{figure}

Figure \ref{fig:OADNC40+DCO40} compares the distribution 
of the DNC and DCO$^+$
emission toward the northern $\int$-shaped filament of the Orion A cloud.
The current mapping is one of the largest single-dish mapping observations of 
deuterated molecules, and
covers a $\sim 5\arcmin\times30\arcmin$ region.
The DCO$^+$ emission down to the 5$\sigma$ sensitivity limit, 
is localized to the northernmost region, OMC-3
(decl. $\sim$ $-4\arcdeg59\arcmin$ to $-5\arcdeg7\arcmin$), whereas 
the DNC emission shows a long filamentary structure spanning  
over OMC-3, 
OMC-2 (decl. $\sim$ $-5\arcdeg7\arcmin$ to $-5\arcdeg17\arcmin$), and 
OMC-1 (decl. $\sim$ $-5\arcdeg17\arcmin$ to $-5\arcdeg28\arcmin$),
down to the region just north of Orion KL
(decl. = $-5\arcdeg22\arcmin30\arcsec$).
For the northern $\int$-shaped filament, 
DCO$^+$ and DNC show different distribution.
\citet{2008PASJ...60..407T} compared the OMC-2 and OMC-3 regions
in N$_2$H$^+$ and H$^{13}$CO$^+$, showing that OMC-2 is 
prominent in N$_2$H$^+$, but OMC-3 is prominent in H$^{13}$CO$^+$
(see their Figures 10 and 11). 
The difference in distribution between DCO$^+$ and DNC may therefore 
partly reflect that between nitrogen-containing and carbon-containing molecules. 
Figure 4 of \citet{1993ApJ...404..643T} also shows that NH$_3$ is bright toward OMC-2
compared with CS.
However, the observed difference could also be influenced by variations in the deuterium fractionation itself, 
rather than simply reflecting the underlying distributions of nitrogen- and carbon-bearing molecules. 
To disentangle these effects, Figure \ref{fig:OADNC40+DCO40} also compares the DNC and DCO$^+$ distributions 
with those of their non-deuterated counterparts, HN$^{13}$C and H$^{13}$CO$^+$, respectively.
We use the $^{13}$C isotopologues rather than HNC and HCO$^+$ 
because the main isotopologue lines are likely optically thick in these regions, 
whereas HN$^{13}$C and H$^{13}$CO$^+$ provide more reliable tracers 
of the column density and remain sufficiently bright for our mapping observations.
The distribution of DNC is not very different from that of HN$^{13}$C, 
but their intensity distribution is rather different.
HN$^{13}$C (color) is strikingly bright toward 
Orion-S (decl. = $-5\arcdeg24\arcmin$).
in the OMC-1 region.
In DNC, OMC-2 is only as bright as OMC-1.
The DNC/HN$^{13}$C column density ratio is a sensitive tracer of 
temperature and chemical evolution, 
decreasing as the gas becomes warmer and more chemically evolved \citep{2010PASJ...62.1473T}.
This behavior is primarily driven by the suppression of deuterium fractionation at elevated temperatures, 
which reduces the formation of DNC through the decline of H$_2$D$^+$ and related deuterated ions, 
while HNC continues to be efficiently produced.
In addition, isotope-exchange reactions and the release of CO back into the gas phase further 
act to decrease the DNC/HN$^{13}$C ratio in warmer, more evolved environments, 
making this ratio particularly high in cold, prestellar gas and lower in star-forming regions.
The fact that the DNC emission shows a filamentary structure down to Orion KL
may suggest that the filament still contains cold gas.
OMC-3 is known as an active intermediate-mass star forming region, 
and contains many molecular outflows
\citep{2000AJ....120.1974Y,2000ApJS..131..465A,2008ApJ...688..344T,2019PASJ...71S...8T}.
Although HCO$^+$ is known to be sensitive to molecular outflow lobes 
\citep{1976ApJ...209..466L,1984ApJ...279..633W}, 
DCO$^+$ does not appear to be significantly affected by them \citep{1997ApJ...487L..93B}.
Moreover, molecular outflows are found not only in OMC-3 but also in OMC-2.
Therefore, the concentration of DCO$^+$ in OMC-3 is difficult to explain solely 
by the activity of molecular outflows.
Instead, the explanation proposed in the preceding paragraph for the difference 
in distribution between DCO$^+$ and DNC is more plausible.

\newpage

Next, we move on to the individual starless cores. Figures \ref{fig:S04-20DNC4} to \ref{fig:S22-32DCO4} illustrate
the distribution of DNC and DCO$^+$ compared with that of HN$^{13}$C (or HNC) and 
H$^{13}$CO$^+$, respectively. 
When the HN$^{13}$C emission is too weak to be clearly displayed, 
we instead show the distribution of HNC.
Core 29 was not detected at the 3$\sigma$ level in either DNC or HN$^{13}$C.
Although small differences are present, the overall distributions of the deuterated molecules broadly resemble those of their non-deuterated counterparts in each core.
This indicates that, while the absolute deuterium fraction may vary from core to core, its spatial distribution within an individual core does not differ strongly from that of the corresponding non-deuterated species.
DNC, HN$^{13}$C and H$^{13}$CO$^+$ spectra toward the core center have already been illustrated in \citet{2020ApJS..249...33K} and \citet{2022ApJ...931...33T}.

\begin{figure}
\includegraphics[bb=0 0 505 825, width=10cm]{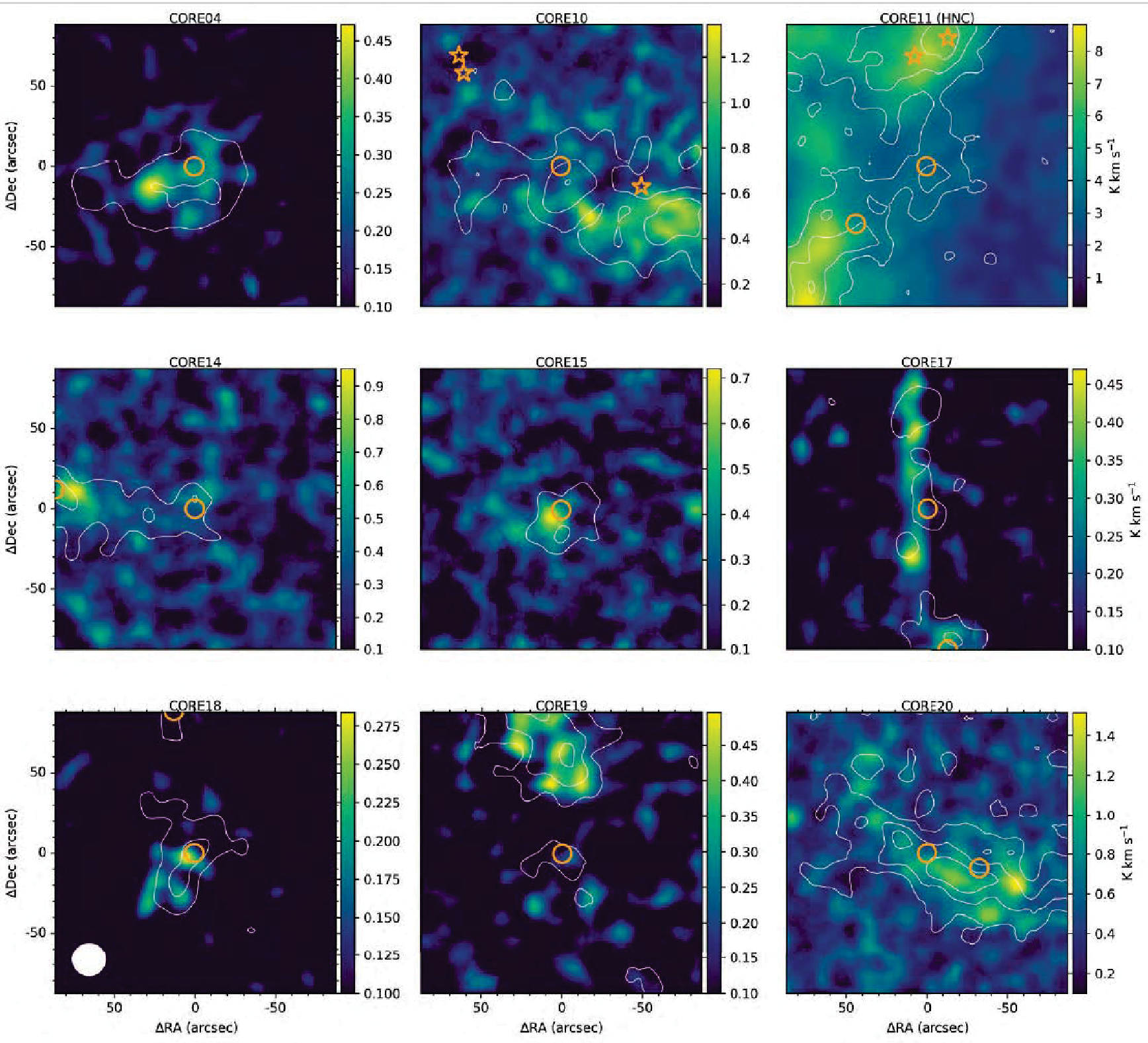}
\caption{
Contours represent the DNC integrated-intensity distribution toward the individual starless cores.
The color shows the integrated-intensity distribution of HN$^{13}$C or HNC for comparison with DNC.
For each core, the velocity integration range was chosen manually to encompass the detected emission.
The contour interval is 0.45 K km s$^{-1}$ ($\sim$3$\sigma$).
The lowest color threshold of 0.1 K km s$^{-1}$ (1.5–3$\sigma$) is used for the HN$^{13}$C and H$^{13}$CO$^+$ 
data from our previous observations. In those observations, the integration times vary among sources to keep the S/N ratio nearly constant.
The orange open circle marks the center (the SCUBA-2 continuum peak) of the starless core \citep{2020ApJS..249...33K}.
The orange star symbol represents \textit{Spitzer} YSOs \citep{2012AJ....144..192M} 
and HOPS (the \textit{Herschel} Orion Protostar Survey) sources \citep{2016ApJS..224....5F}.
Offsets are given relative to the adopted map center.
The white filled circle represents the HPBW for the deuterated molecules.
\label{fig:S04-20DNC4}}
\end{figure}

\newpage

\begin{figure}
\figurenum{2}
\includegraphics[bb=0 0 505 825, width=10cm]{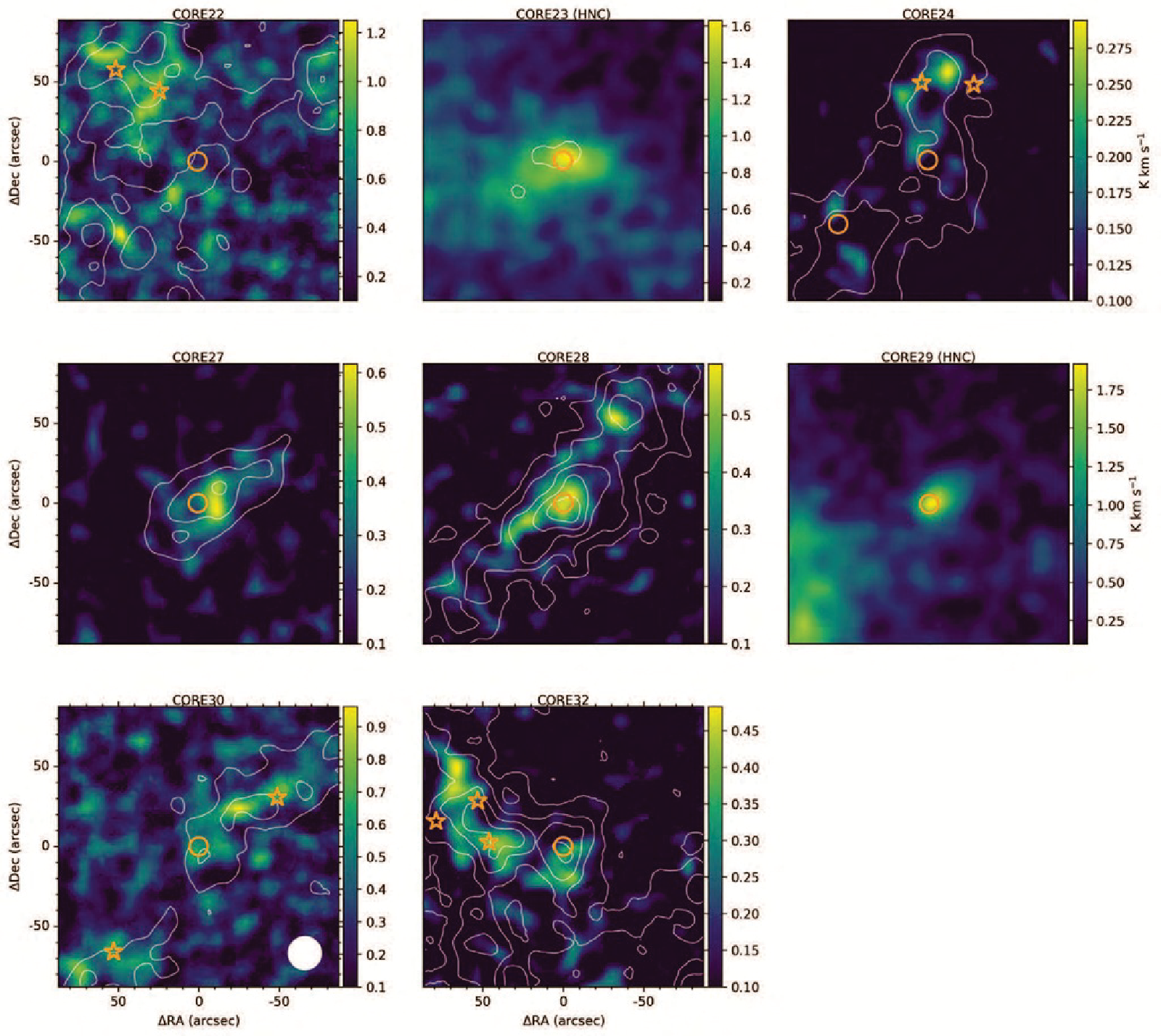}
\caption{
Continued. \label{fig:S22-32DNC4}}
\end{figure}

\newpage

\begin{figure}
\includegraphics[bb=0 0 505 825, width=10cm]{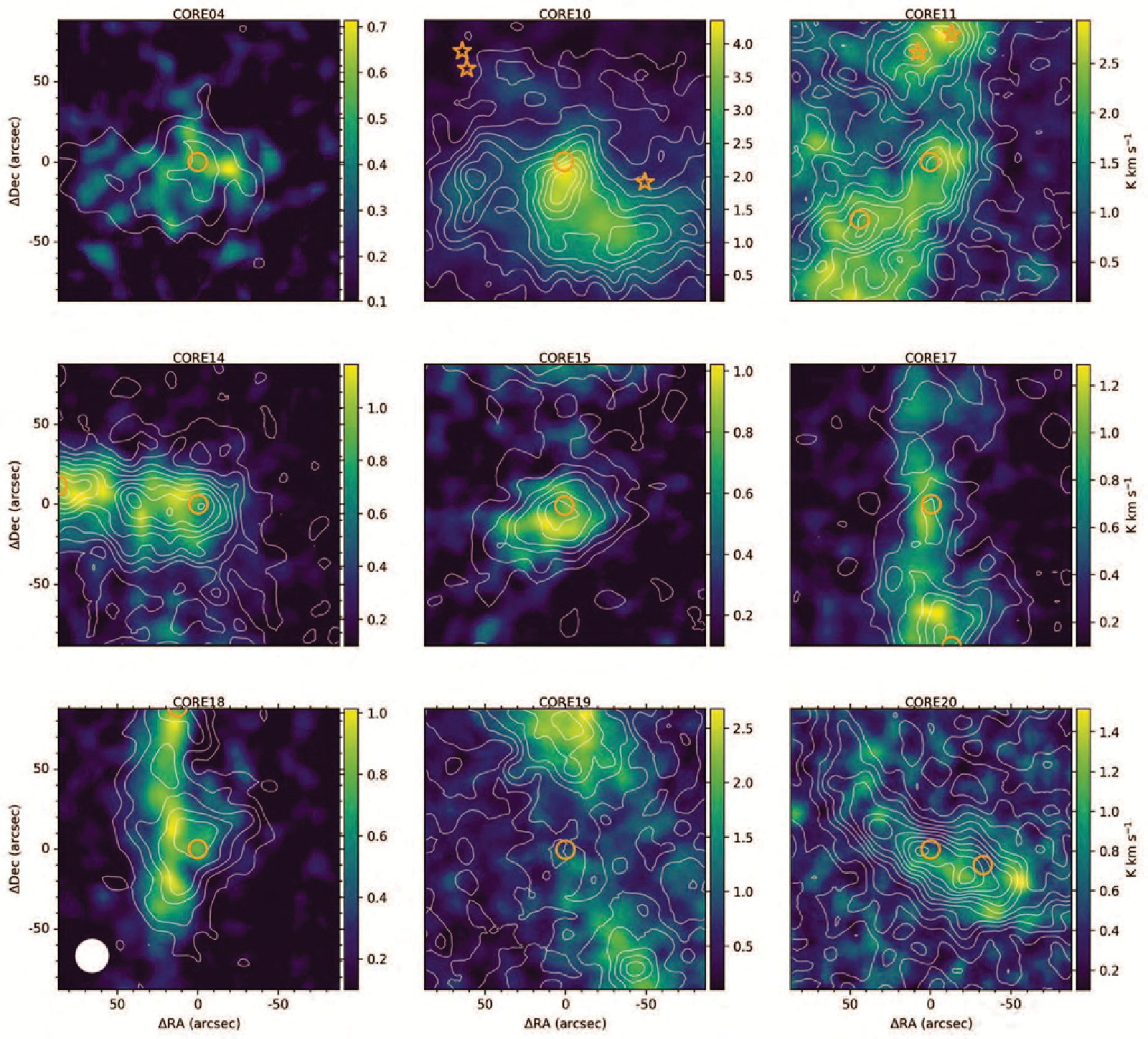}
\caption{
Contours represent the DCO$^+$ integrated intensity toward the individual starless cores..
The contour interval for DCO$^+$ is 0.45 K km s$^{-1}$ (3$\sigma$).
The color shows the H$^{13}$CO$^+$ integrated intensity for comparison with DCO$^+$.
The white filled circle represents the HPBW for the deuterated molecules.
\label{fig:S04-20DCO4}}
\end{figure}
\newpage

\begin{figure}
\figurenum{3}
\includegraphics[bb=0 0 505 825, width=10cm]{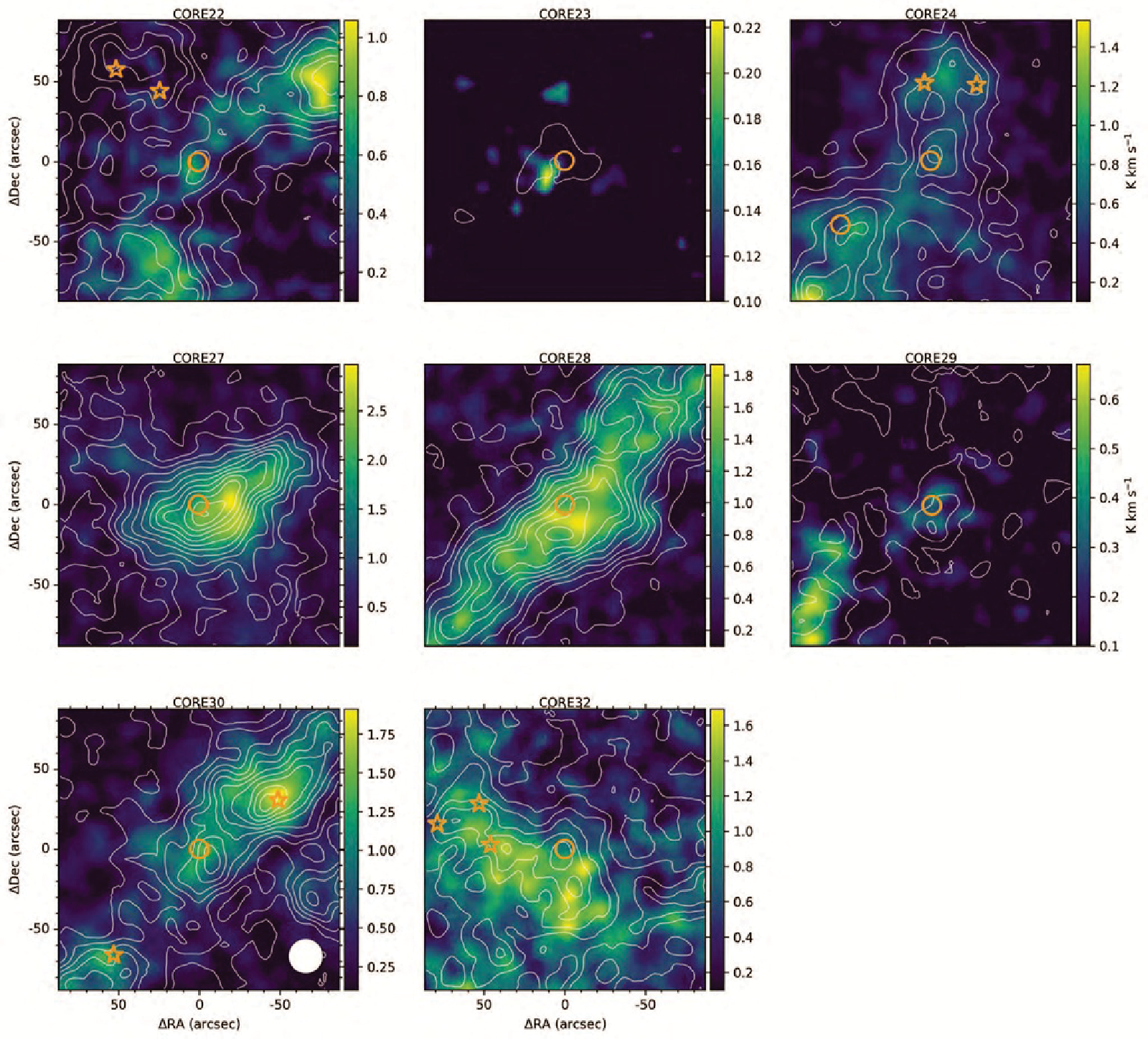}
\caption{Continued. \label{fig:S22-32DCO4}}
\end{figure}
\newpage

\subsection{Column Density Ratio}

The integrated-intensity ratio maps 
of the deuterated molecules to the non-deuterated molecules
($W$ (DNC)/$W$ (HN$^{13}$C) and $W$ (DCO$^+$)/$W$ (H$^{13}$CO$^+$)
of the individual starless cores often show 
that the core centers tend to have lower ratios
than the core edges.
Such a tendency is expected 
when the line optical depth of the isotopic non-deuterated molecules is
negligible but that of the deuterated molecules is not small.
According to \citet{2001ApJ...547..814H},
DNC can be optically thick toward dark cloud cores.
To examine whether this effect is present in our data, 
we selected five relatively isolated starless cores (04, 15, 18, 27, and 28), 
where the cores are well defined and minimally affected by confusion from neighboring structures.
For these cores, the average line-center optical depth of DNC is estimated to be 1.3.
We therefore adopt the column density ratio maps rather than 
the integrated-intensity ratio maps in the following analysis.

The column density is calculated by using the standard 
LTE (local thermodynamic equilibrium)
method. We followed the formulation by \citet{1992ApJ...392..551S}.
We calculate the FWHM (full width at half maximum) linewidth $\Delta v$
from the ratio of the velocity-integrated intensity
$\int T_R~dv$ to the peak intensity $T_R~ (max)$: 
$\Delta v = 2 \sqrt{ln~2/\pi}  \int T_R~dv / T_R~ (max)$ for the Gaussian profile.
The line optical depth $\tau$ is calculated using the peak intensity
$T_R (max)$,  the excitation temperature $T_{ex}$, and the temperature
of the cosmic microwave background radiation $T_{bg}$ (2.725 K), by using the task
``comb'' in the software AIPS:
$T_R (max) = [J_{\nu}(T_{ex})-J_{\nu}(T_{bg})][1-$exp$~(-\tau)],$
where $J_{\nu}(T) = (h\nu/k)~[$exp$~(h\nu/kT)-1]^{-1}$.
$T_{ex}$ is assumed to be half the gas kinetic temperature
($T_{ex} = 0.5~T_k$)
\citep{1992ApJ...392..551S,2008PASJ...60..407T,2017ApJS..228...12T},
or the levels were assumed  50\% subthermally excited.
The critical densities of N$_2$H$^+$, HNC, HCO$^+$, and H$^{13}$CO$^+$ are 
$(6-14)\times10^4$ cm$^{-3}$ \citep[][and references therein]{2022ApJ...931...33T}, 
and those of DNC, DCO$^+$, and HN$^{13}$C
will not be very different.
Then, we assume that all the $J = 1-0$ transitions we observed are 
similarly subthermal.
We assume constant gas kinetic temperatures $T_k$
of 20 K and 10$-$20 K for the northern $\int$-shaped filament 
\citep{2008PASJ...60..407T} and 
for the individual starless cores
\citep{2010A&A...524A..91M,2014PASJ...66...16T,2014ApJ...789...83T,2017ApJS..228...12T}, respectively.
$T_{ex}$ for N$_2$H$^+$ toward the $\int$-shaped filament is estimated to be 
$\sim$10 K
\citep{2008PASJ...60..407T}.
We assume $T_{ex}$ = 5$-$10 K for the individual starless cores,
and investigate here two extreme cases of Tex = 5 and 10~K.
For DCO$^+$ and H$^{13}$CO$^+$, $T_{ex}$ = 5~K 
cannot reproduce the observed intensity in general, 
and therefore we assume $T_{ex}$ = 10~K (thermalized) for
these molecules toward the individual cores.
For the starless cores, we will present the $N$~(DNC)/$N$~(HN$^{13}$C) maps with $T_{ex}$ = 10~K in the main text,
whereas the $N$~(DNC)/$N$~(HN$^{13}$C) maps with $T_{ex}$ = 5~K
and the $N$~(DCO$^+$)/$N$~(H$^{13}$CO$^+$) maps with $T_{ex}$ = 10~K
will be presented in the Appendix,
for readability.
Even if $T_{ex}$ is increased to 15–20 K, the derived deuterium fractions change only slightly, 
and the conclusions of this paper remain unaffected. However, if $T_{ex}$ differs significantly 
between the two molecules, the derived deuterium fractions would change accordingly.
The thresholds for the column density calculation are the integrated-intensity levels of 0.3, 0.5, 0.6, and 0.8~K~km~s$^{-1}$ for DNC, HN$^{13}$C, DCO$^+$, and H$^{13}$CO$^+$, respectively, toward the northern $\int$-shaped filament.
For all lines toward the individual starless cores, the threshold is 0.1~K~km~s$^{-1}$.

\subsection{Column Density Ratio along the Northern $\int$-shaped Filament}

\begin{figure}
\includegraphics[bb=0 0 350 450, width=15cm]{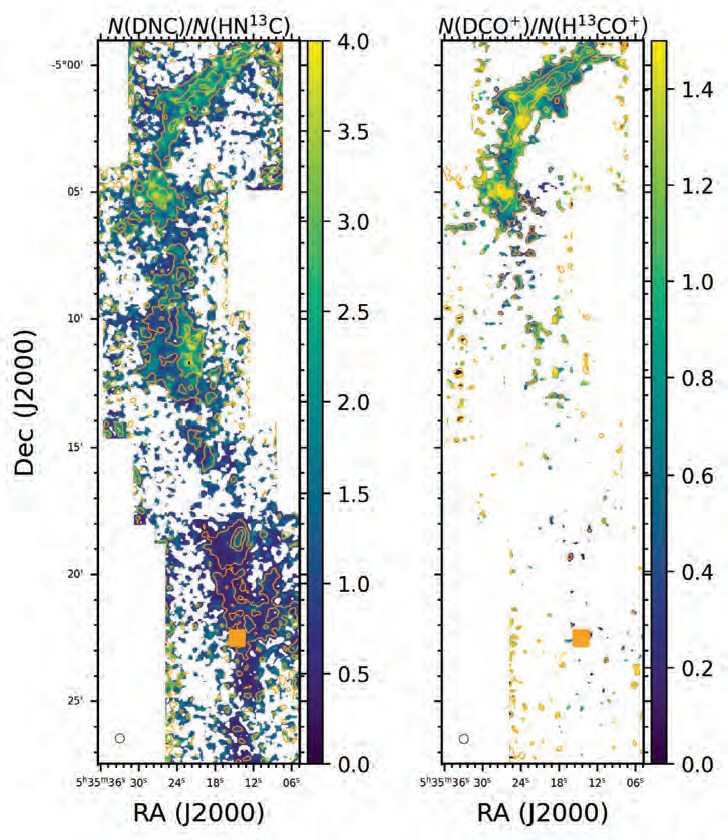}
\caption{
The color shows the column density ratios, $N$(DNC)/$N$(HN$^{13}$C) (left) and $N$(DCO$^+$)/$N$(H$^{13}$CO$^+$) (right), 
toward the northern $\int$-shaped filament.
Contours represent the integrated-intensity distributions of DNC (left) and DCO$^+$ (right).
The contour interval is 1 K km s$^{-1}$ ($\sim$5$\sigma$).
The open circle in the lower-left corner represents the HPBW for the deuterated molecules.
\label{fig:ZOADNCHN140+DCOH140}}
\end{figure}

Figures \ref{fig:ZOADNCHN140+DCOH140}
shows the column density ratio maps
of $N$~(DNC)/$N$~(HN$^{13}$C) 
and 
$N$~(DCO$^+$)/$N$ (H$^{13}$CO$^+$)
toward the northern $\int$-shaped filament.
The column density ratio is calculated by ignoring the difference in the HPBW 
(21$\farcs$4 and 18$\farcs$4 for the deuterated and non-deuterated molecules, respectively), 
corresponding to differences of 16\% in HPBW and 35\% in beam area.
The D/H column density ratio does not show large variation (more than a few) 
on scales of 1 arcmin (0.13 pc).
This result is consistent with those by \citet{2003ApJ...594..859H} and \citet{2019A&A...629A..15R}.
$N$~(DNC)/$N$~(HN$^{13}$C) seems to decrease from north to south
along the northern $\int$-shaped filament.
Such a north–south gradient is consistent with the picture that regions 
with higher deuterium fraction represent an earlier evolutionary stage 
characterized by strong CO depletion and 
centrally concentrated density structures \cite{2005ApJ...619..379C}, 
whereas the lower ratios toward OMC-1 indicate more chemically evolved and warmer conditions.

\begin{figure}
\includegraphics[bb=0 0 505 475, width=10cm]{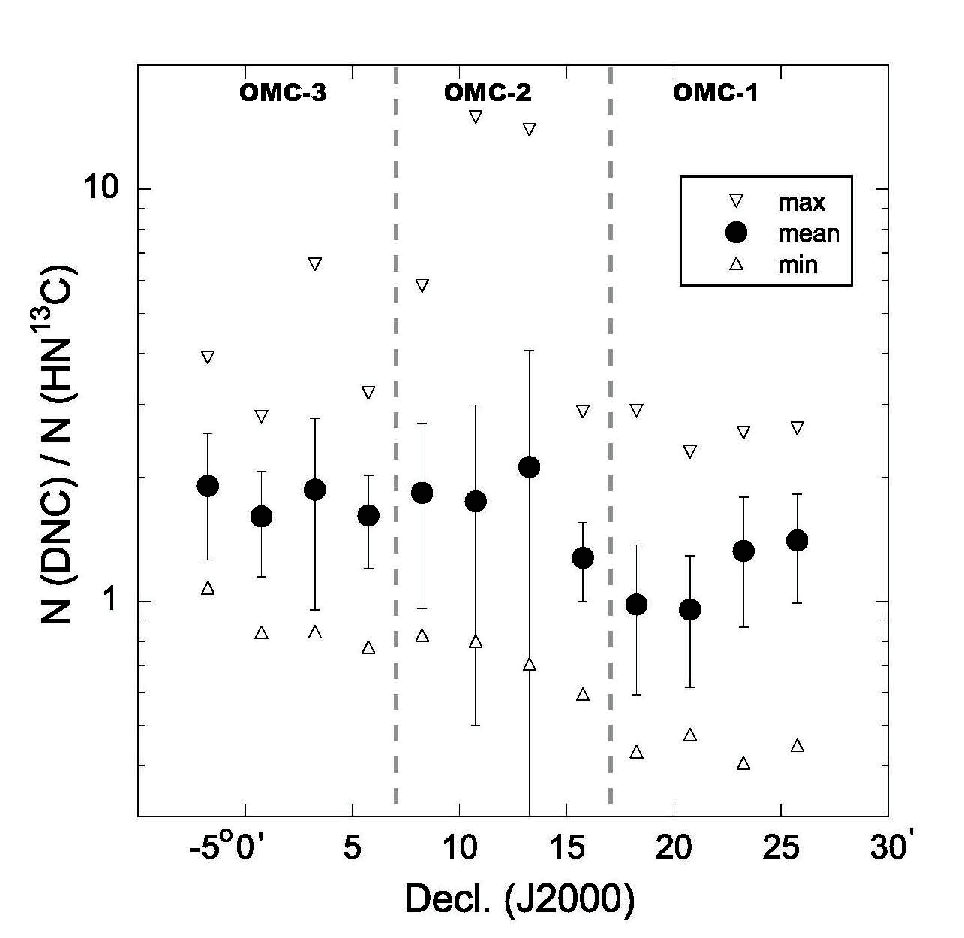}
\caption{The column density ratio
$N$~(DNC) /$N$~(HN$^{13}$C)
along declination.
The statistics were calculated for each $150\arcsec$-wide strip in declination.
The horizontal gray dashed lines mark the boundaries between OMC-1, OMC-2, and OMC-3.
The `error bars' represent the standard deviation of the data within each strip, which has 
a length of $\Delta$R.A. $\sim 5\arcmin$, rather than the measurement errors.
\label{fig:DNC-ratio-Decl}}
\end{figure}

Figure \ref{fig:DNC-ratio-Decl} illustrates the DNC/HN$^{13}$C
column density ratio along declination (each 150$\arcsec$-wide strip).
The ratio shows high values in OMC-2 and OMC-3, 
and low values in OMC-1.
It seems that the OMC-2 and OMC-3 regions
show high ($\sim 2$) deuterium fraction gas, suggesting that
it still contains molecular gas close to
the onset of star formation \citep{2006ApJ...646..258H}.
The column density ratio of DNC to HN$^{13}$C is low around the Ori KL region 
(decl. $\sim$ $-5\arcdeg22\arcmin30\arcsec$), 
possibly due to the higher gas temperature known in this region
\citep{2010PASJ...62.1473T}. 
Because the DCO$^+$ emission is detected only in the OMC-3 region, we cannot
discuss the $N$~(DCO$^+$)/$N$ (H$^{13}$CO$^+$) variation along the filament.
The scatter in the `error bars' is relatively large because each strip ($\Delta$R.A. $\sim 5\arcmin$) 
spans the filament and therefore includes variations in the deuterium fractionation on scales 
larger than $\sim 1\arcmin$.
Nevertheless, the average values do not vary significantly 
within the OMC-2 and OMC-3 regions compared with the `error bars', 
and they are systematically larger than those in the OMC-1 region.

\citet{2024A&A...687A..70S} reported the distribution of 
HCO$^+$ ($J$ = 1$-$0) 
and DCO$^+$ ($J$ = 3$-$2) 
across the entire $\int$-shaped filament 
using the IRAM 30 m telescope. 
DCO$^+$ ($J$ = 3$-$2) is weak or undetected toward OMC-2,
which is similar to our DCO$^+$ ($J$ = 1$-$0) result.
DCO$^+$ ($J$ = 3$-$2) is brightest toward Orion KL in OMC-1,
which is very different from our DCO$^+$ ($J$ = 1$-$0) result.
This is probably due to difference in the upper state energy level
between these lines.
Although the transitions are different and 
a direct quantitative comparison is not possible, their conclusion that the 
DCO$^+$/HCO$^+$ ratio is enhanced in OMC-3 is fully consistent with our results. 
Our result based on the optically thinner line of H$^{13}$CO$^+$
and similar upper energy levels strengthens the conclusion.
The higher abundance of DCO$^+$ in OMC-3 is interpreted by moderately heavy CO depletion.  
In general, CO depletion becomes weak toward more evolved cores 
where CO has returned to the gas phase
\citep{2019A&A...629A..15R,2024A&A...688A.118L}.
Our results for DNC and DCO$^+$ also follow the evolutionary trend established for 
N$_2$D$^+$/N$_2$H$^+$, where higher deuterium fraction traces cold, dense, 
and dynamically young gas, while lower ratios indicate 
chemically evolved conditions 
\citep{2005ApJ...619..379C,2019PASJ...71...73T}.

\newpage

\newpage

\subsection{Column Density Ratio in the Individual Starless Core}

Figures \ref{fig:ZS04-20DNC_10} and \ref{fig:ZS22-32DNC_10}
show the column density ratio maps
of $N$~(DNC)/$N$~(HN$^{13}$C) with $T_{ex}$ = 10 K
toward the individual starless cores.
Some DNC/HN${13}$C column density ratio map show 
'white areas' with no ratio values near the core center,
because the assumed excitation temperature is too low to
reproduce the observed intensity (e.g., core 28)
or one of the two relevant molecules is undetected. 
%Core 29
%has the color only near the core center, because 
%the HN$^{13}$C emission was detectable only there.

The column density ratio maps of DNC to HN$^{13}$C with $T_{ex}$ = 5 K are similar
to those with $T_{ex}$ = 10 K, and are shown in the Appendix,
together with
the $N$~(DCO$^+$)/$N$ (H$^{13}$CO$^+$) maps assuming $T_{ex}$ = 10 K. 
Given that the excitation temperature is
higher than the cosmic background radiation temperature,
the column density is approximately proportional to the integrated
intensity, and the ratio of those of two molecules will be 
less sensitive to the excitation temperature.
We hence believe that the derived column density ratio should be reliable,
given that the two molecules have similar critical densities and
excitation temperatures.
The column density ratio is rather constant within each starless core map.
This result is again consistent with that by \citet{2003ApJ...594..859H}.
Cores 17, 18, 23, and 27 may show slight enhancement in 
$N$~(DNC)/$N$ (HN$^{13}$C), but the evidence is not conclusive.
The distribution of the ratio
$N$~(DCO$^+$)/$N$ (H$^{13}$CO$^+$) in the appendix 
is nearly flat for all the starless cores except for core 4.

\begin{figure}
\includegraphics[bb=0 0 505 825, width=10cm]{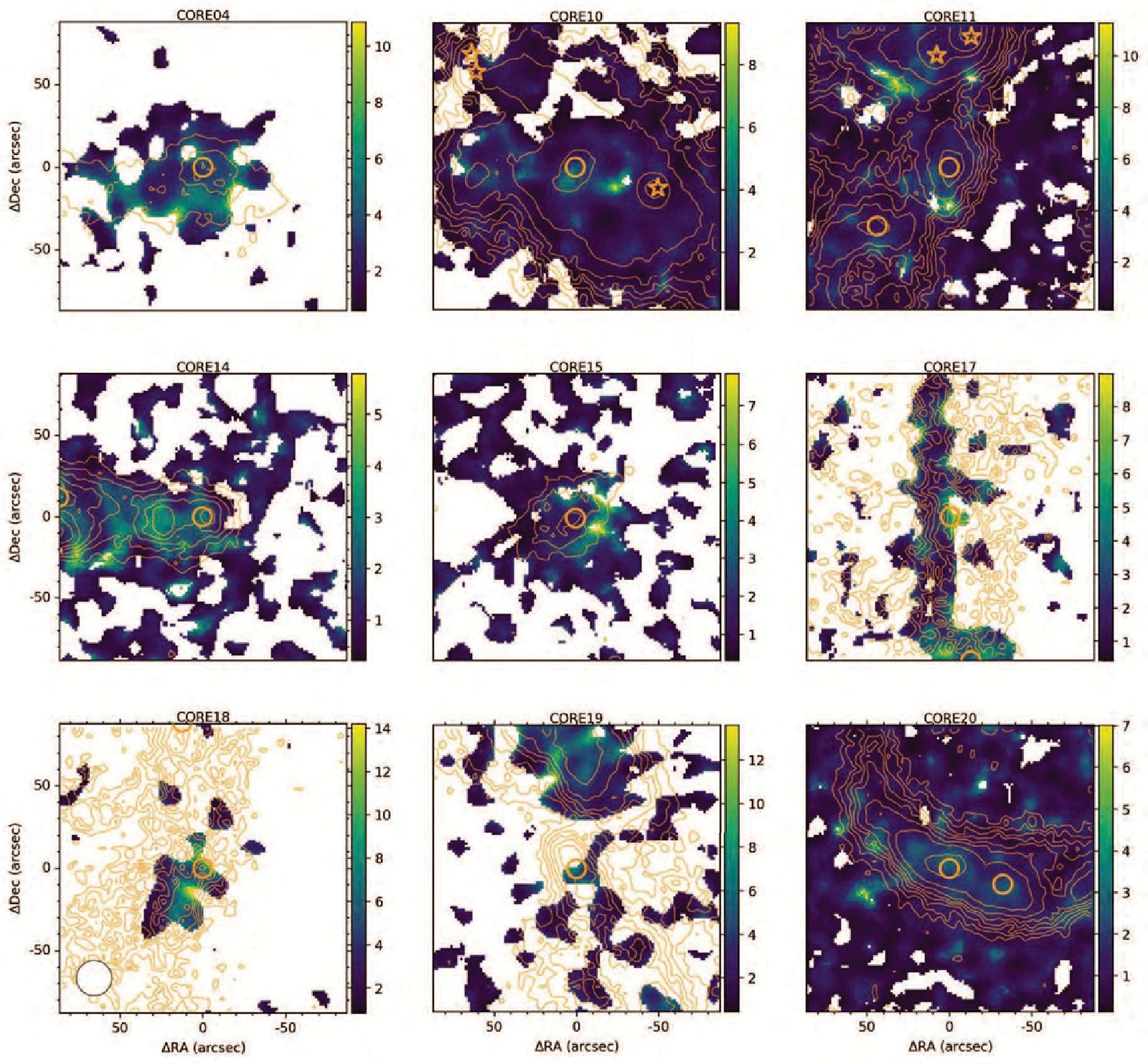}
\caption{The color shows the column density ratio $N$(DNC)/$N$(HN$^{13}$C) toward the 
individual starless cores, calculated assuming $T_{\rm ex}$ = 10 K.
Contours represent the 850 $\mu$m continuum distribution from \citet{2018ApJS..236...51Y}; 
the contour levels are 30, 60, 90, 120, 150, 200, 300, 600, and 1000 mJy beam$^{-1}$.
The open circle represents the HPBW for the deuterated molecules.
\label{fig:ZS04-20DNC_10}}
\end{figure}
\newpage

\begin{figure}
\figurenum{6}
\includegraphics[bb=0 0 505 825, width=10cm]{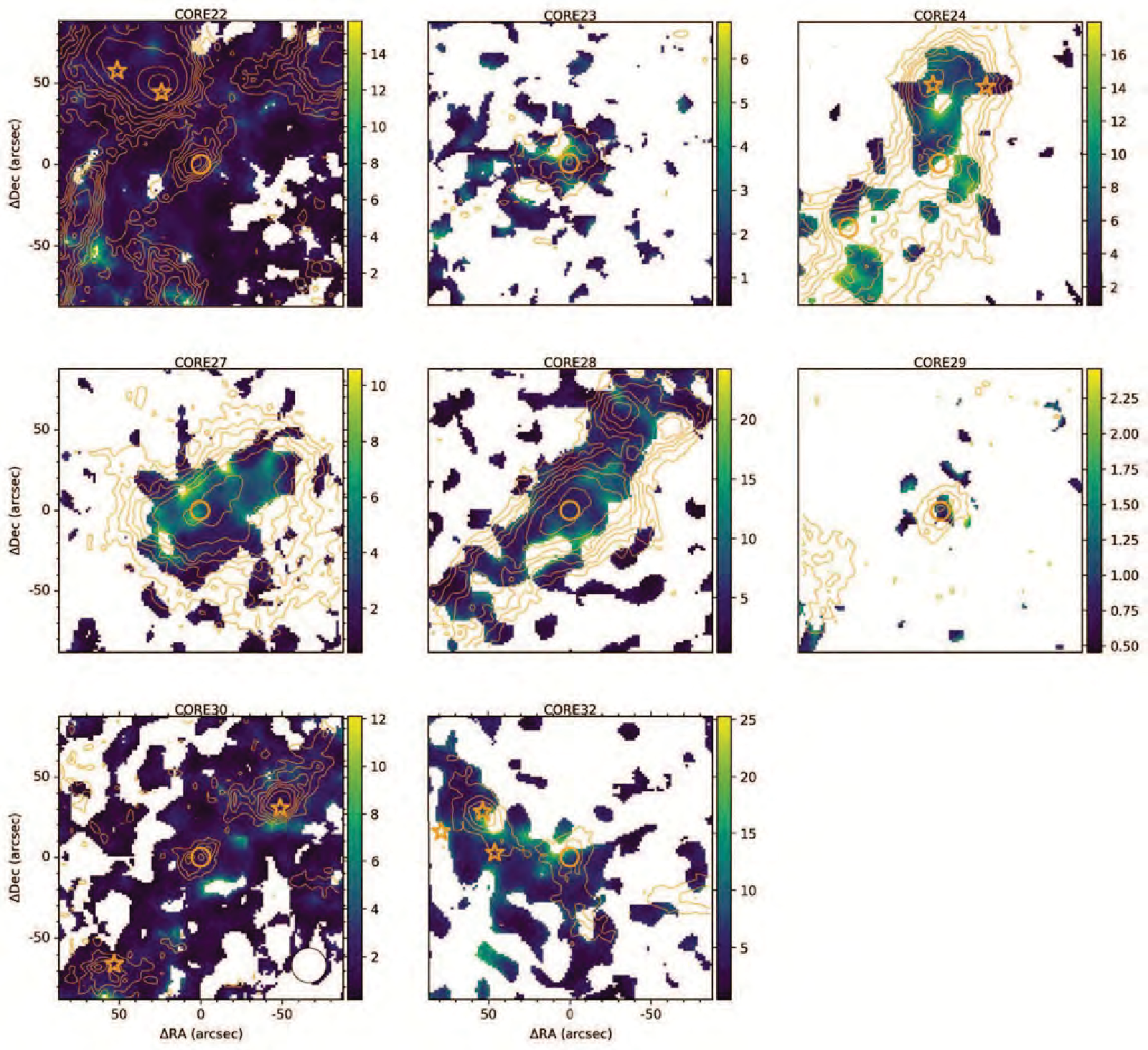}
\caption{Continued. \label{fig:ZS22-32DNC_10}}
\end{figure}
\newpage

\subsection{Radial Variation of the Column Density Ratio within the Core}

We are interested in variation of deuterium fraction within a molecular cloud core.
Deuterium fraction can be described as the ratio of the column density
of the deuterated molecule to that of the corresponding non-deuterated molecule.
\cite{2005ApJ...619..379C} showed that
starless cores with higher deuterium fraction in the N$_2$D$^+$/N$_2$H$^+$
column density ratio tend to have larger CO depletion,
more compact radial density profile, and higher central H$_2$ column density.
\citet{2003ApJ...594..859H}
observed nearby dark cloud cores, TMC-1, L1512, L1544, and L63 in Taurus,
and found that deuterium fraction of DNC/HN$^{13}$C
ratio is rather constant within each core,
but the core-to-core variation ranges over an order of magnitude.
On the other hand,
\citet{2019A&A...629A..15R} 
showed that deuterium fraction of N$_2$D$^+$/N$_2$H$^+$ decreases outward from the core center
while that in HCO$^+$ is rather constant in L1544 of Taurus.

The degree of the deuterium fraction differs among molecules 
even if we observe the same core.
\citet{2017ApJS..228...12T} and \citet{2020ApJS..249...33K} introduced 
the chemical evolution factor (CEF)
in terms of the average deuterium fraction of a few deuterated molecules 
with empirical scaling factors
from their own observations toward Orion 
and from the literature on the nearby cold molecular cloud core.
Their Orion observations were made in single-pointing
toward the core center
identified with the JCMT and SCUBA-2 camera at 850 $\mu$m 
\citep{2018ApJS..236...51Y}.
In general, high density starless cores seem to have higher deuterium fractions
in comparison among cores \citep{2005ApJ...619..379C}, 
and one may expect that the core center has the highest deuterium fraction 
(within a core) if the core has a radial density gradient decreasing outward from the center.

\begin{figure}
\includegraphics[bb=0 0 505 625, width=10cm]{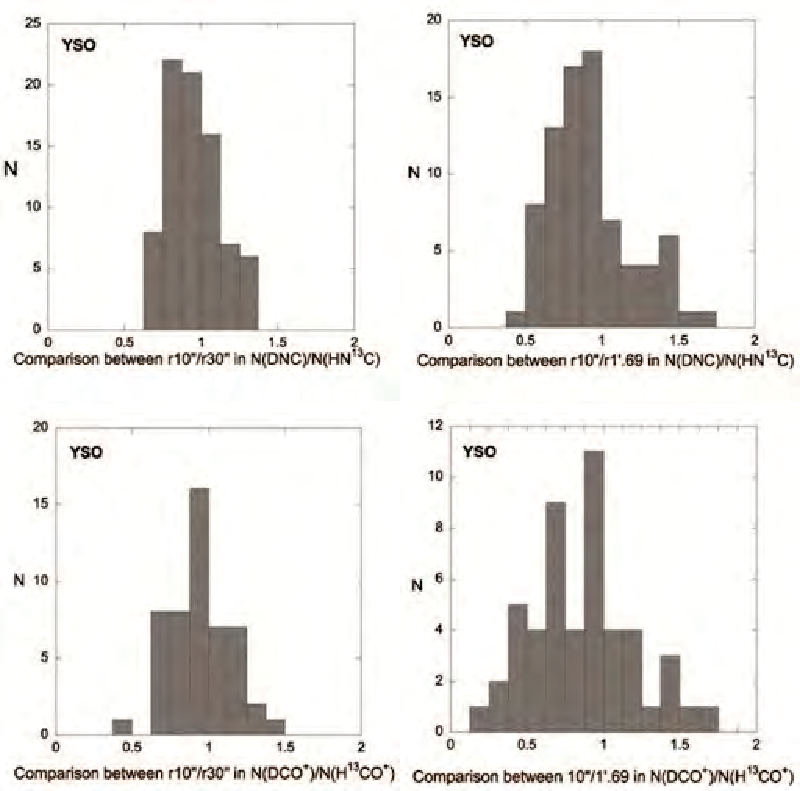}
\caption{
Histograms of the comparisons of different measurement radii 
of the column density ratios toward the YSOs.
\label{fig:r-ratio-stars}}
\end{figure}

\begin{figure}
\includegraphics[bb=0 50 505 850, width=10cm]{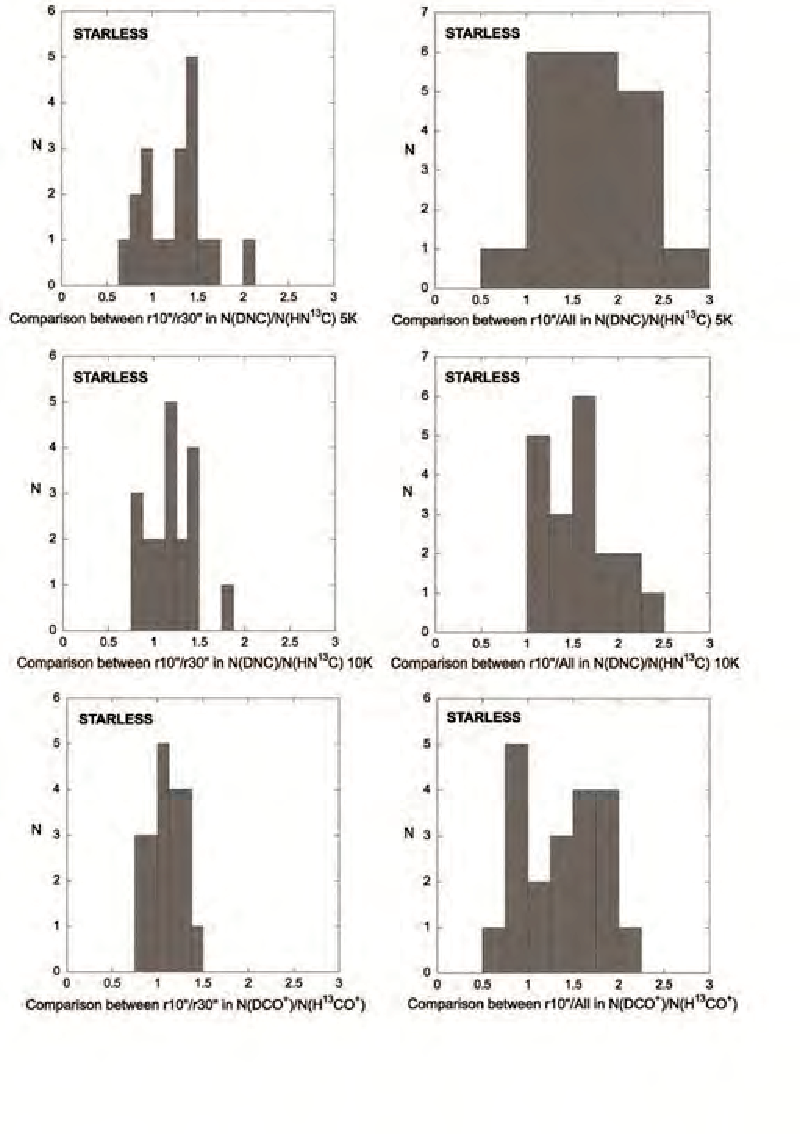}
\caption{
Histograms of the comparisons of different measurement radii
of the column density ratios toward the starless cores.
\label{fig:r-ratio-starless}}
\end{figure}

Figures \ref{fig:r-ratio-stars} and \ref{fig:r-ratio-starless} 
illustrate histograms comparing 
the column density ratios using different measurement radii from the core center,
for the northern $\int$-shaped filament and the individual starless cores, respectively.
$r = 10\arcsec$ is very close to the telescope beam radius (half of the HPBW), 
$r = 30\arcsec$ is three times the telescope beam radius.
'All' for the individual starless cores are the average over each map.
$r = 1\farcm69$ for the northern $\int$-shaped filament
is chosen to be equivalent to 
the map size (3$\arcmin$ square) of the individual starless core.
If the column density peaks at the core center, the peak in the histogram should be
larger than unity.
The ratio of $r = 10\arcsec$ to $r = 30\arcsec$
mostly ranges from 0.8 and 1.5,
and that of $r = 10\arcsec$ to the core map area
mostly ranges from 0.8 to 2.5.
Then, we conclude that the column density ratio
may show a slight increase toward the core center, but 
the difference between the center and the core map area average
is typically a factor of 2$-$3 at most.
On the other hand, $N$~(DNC)/$N$~(HN$^{13}$C) at the core center
is 4.30$\pm$3.24, and the variation is about a factor of several.
It sems that the radial variation within each core is smaller than the core-to-core variation.

Let us go into details.
Figure \ref{fig:r-ratio-starless} shows that 
the ratio of $r = 10\arcsec$ to the core map area
is larger than unity, suggesting that the core center
tends to have a slightly (a factor of 1$-$2) larger deuterium fraction
than the core area, for the starless cores.
Figures \ref{fig:r-ratio-stars} shows that 
the ratio of $r = 10\arcsec$ to the core map area
has a distribution peak below unity for $N$~(DNC)/$N$~(HN$^{13}$C), 
suggesting
that the YSO position
has slightly (a factor of 1$-$2) lower deuterium fraction.
This tendency is not clear for 
$N$~(DCO$^+$)/$N$~(H$^{13}$CO$^+$).
Figures \ref{fig:r-ratio-stars} and \ref{fig:r-ratio-starless} show that
the ratio of $r = 10\arcsec$ to the core map area
has larger deviation than that
of $r = 10\arcsec$ to $r = 30\arcsec$.
It may suggest that even inside the core map area,
a slight chemical variation starts to appear.

The D/H column density ratio shows little variation within
molecular cloud cores (typical diameter 0.1 pc).
The molecular cloud cores therefore appear to be chemically relaxed.
\citet{2003ApJ...594..859H} found that the
DNC/HN$^{13}$C ratio is nearly constant within each core
and explained this result as follows.
The typical timescale of deuterium fractionation is of order $10^5$ yr
for a density of $10^5$ cm$^{-3}$ \citep{2001ApJS..136..579T}.
The dynamical timescale, or the free-fall time, is also of order $10^5$ yr
for a density of $10^5$ cm$^{-3}$.
When gas-phase formation and desorption are taken into account,
the effective depletion timescale is also of order $10^5$ yr
in molecular cloud cores \citep{2001ApJ...552..639A,2003ApJ...593..906A}.
Thus, all of these timescales are similar in magnitude.
The same explanation could also account for our results.
The turbulent crossing time across a typical core diameter,
assuming an FWHM linewidth of 0.5 km s$^{-1}$ \citep{2020ApJS..249...33K},
is $2\times10^5$ yr, which is comparable in magnitude.

\citet{2019PASJ...71...73T} mapped the N$_2$D$^+$ emission 
over $2\arcmin\times2\arcmin$ region toward Tau MC5N  by using the Nobeyama 45~m telescope,
and discussed high deuterium fraction of N$_2$D$^+$/N$_2$H$^+$ toward
a centrally concentrated
starless condensation.
This result suggests that the depletion of CO is greater than that of N$_2$.
Regarding H$_2$D$^+$, which is the parent molecule for deuterated molecules,  
\citet{2025ApJ...992...55T} reported that
the ortho-H$_2$D$^+$ distribution becomes compact only within a few $\times10^4$
years immediately before or after protostar formation.
\citet{2019A&A...629A..15R}
mapped
L1544 ($2\arcmin\times2\arcmin$ area) in N$_2$D$^+$ and DCO$^+$
using the IRAM 30~m telescope and obtained the evidence of N$_2$D$^+$ depletion.
They showed that the N$_2$D$^+$/N$_2$H$^+$ column density ratio
decreases outward from the core center in L1544, but
only by a factor of 2$–$3.

\newpage

\newpage

\subsection{Relationship with the Inward Motions}

\cite{2005ApJ...619..379C} showed that
starless cores with higher deuterium fraction in the N$_2$D$^+$/N$_2$H$^+$
column density ratio tend to have broader N$_2$H$^+$ line with asymmetry 
suggesting inward motions.
\citet{2013ApJ...777..121S} claimed that 
HCO$^+$ ($J$ = 3$-$2) inward motions were observed toward high D/H starless cores defined in 
N$_2$D$^+$/N$_2$H$^+$.
\citet{2005ApJ...619..379C} obtained similar result on the basis of the
N$_2$H$^+$ linewidth and line shape.
Infall motions may be better traced in HCO$^+$ than in N$_2$H$^+$,
because of its larger optical depth.
In the Orion regions,
\citet{2022ApJ...931...33T} reported that cores 10, 11, and 14 to have the 
HCO$^+$ ($J$ = 1$-$0) blue-skewed
line profiles, whereas cores 18 and 32 were identified as
candidate blue-skewed line profiles.
Table \ref{tbl:inward} compares the deuterium fraction for
the three datasets,
(a) all the starless cores, 
(b) the blue-skewed cores (10, 11, and 14), 
(c) the blue-skewed and candidate cores (10, 11, 14, 18, and 32).
We do not find evidence that the cores with
the HCO$^+$ ($J$ = 1$-$0)  blue-skewed profiles
have high values of $N$~(DNC)/$N$~(HN$^{13}$C) or  $N$~(DCO$^+$)/$N$~(H$^{13}$CO$^+$).
Categories (b) and (c) show the largest ratio of $\sim$2 in $N$~(DNC)/$N$~(HN$^{13}$C) at 5 K,
when we compare $r = 10\arcsec$ and all.

\floattable
\begin{deluxetable}{lccccccccc}
\rotate
\tablecaption{Deuterium Fraction toward the Starless Cores with the Inward Motions Signature \label{tbl:inward}}
\tablecolumns{9}
\tablenum{1}
\tablewidth{0pt}
\tablehead{
\colhead{Category} &
\multicolumn{6}{c}{$N$~(DNC)/$N$~(HN$^{13}$C)} &
\multicolumn{3}{c}{$N$~(DCO$^+$)/$N$~(H$^{13}$CO$^+$)} 
\\
\cline{2-10}
\colhead{} &
\multicolumn{3}{c}{$T_{ex}$ = 5 K} &
\multicolumn{3}{c}{$T_{ex}$ = 10 K} &
\multicolumn{3}{c}{$T_{ex}$ = 10 K} 
\\
\cline{2-10}
\colhead{} &
\colhead{$r 10\arcsec$} &
\colhead{$r 30\arcsec$} &
\colhead{All} &
\colhead{$r 10\arcsec$} &
\colhead{$r 30\arcsec$} &
\colhead{All} &
\colhead{$r 10\arcsec$} &
\colhead{$r 30\arcsec$} &
\colhead{All} 
\\
\colhead{} &
\colhead{} &
\colhead{} &
\colhead{} &
\colhead{} &
\colhead{} &
\colhead{} &
\colhead{} &
\colhead{} 
}
\startdata
(a) all	&	4.90 	$\pm$	3.64 	&	4.07 	$\pm$	3.04 	&	3.03 	$\pm$	2.34 	&	4.29 	$\pm$	2.73 	&	3.73 	$\pm$	2.56 	&	2.79 	$\pm$	2.07 	&	4.55 	$\pm$	2.10 	&	4.05 	$\pm$	1.67 	&	3.26 	$\pm$	1.31 	\\
(b) inward	&	2.71 	$\pm$	0.23 	&	2.10 	$\pm$	0.11 	&	1.33 	$\pm$	0.17 	&	2.45 	$\pm$	0.17 	&	1.97 	$\pm$	0.19 	&	1.30 	$\pm$	0.16 	&	4.24 	$\pm$	3.99 	&	3.29 	$\pm$	2.39 	&	2.58 	$\pm$	1.37 	\\
(c) inward+candidate	&	5.95 	$\pm$	4.75 	&	4.17 	$\pm$	2.87 	&	2.88 	$\pm$	2.16 	&	4.81 	$\pm$	3.38 	&	3.81 	$\pm$	2.59 	&	2.75 	$\pm$	2.06 	&	4.84 	$\pm$	3.00 	&	4.01 	$\pm$	2.14 	&	2.82 	$\pm$	1.06 	\\
\enddata
\end{deluxetable}

\subsection{Implications for the ALMA Observations}

The DNC and  HN$^{13}$C intensity distribution, and 
$N$~(DNC)/$N$~(HN$^{13}$C) distribution in the starless cores provide us
with some implications for ALMA observations.
It is difficult to detect nearby  ($<$ 500 pc) starless cores 
with the ALMA 12m Array,
but possible to image them with the ALMA-ACA
\citep{2016ApJ...823..160D,2020ApJ...899...10T}.
It is most likely due to the size scale of the density structure, 
which results in the missing flux.
\citet{2025ApJ...992...55T} and \citet{2025ApJ...995...36L} detected the H$_2$D$^+$
emission toward nearby starless cores with the ALMA-ACA.
%\citet{2015ApJ...803...70S} discussed that the distribution of DNC and 
%HN$^{13}$C
%are appreciably different in ALMA observations toward an IRDC (infrared dark cloud).
To study the evolution of molecular cloud cores, 
both starless and star-forming, 
in nearby star-forming regions 
such as Taurus and Orion 
through the deuterium fraction,
it is important to include the ACA Array or relevant single-dish data.
Different degree in the missing flux between 
the deuterated and non-deuterated molecules 
should affect the deuterium fraction estimate seriously.

%RADEX
%\citet{2007A&A...468..627V}

%LAMDA
%Leiden Atomic and Molecular Database
%\citet{2005A&A...432..369S}

\section{SUMMARY \label{sec:sum}}

We observed the northern $\int$-shaped filament
and the 20 individual starless cores in the Orion A and B clouds
in DNC and DCO$^+$ with 7BEE and
the Nobeyama 45~m radio telescope.
For the northern $\int$-shaped filament,
we found that the DNC emission is distributed along the filament, 
whereas the DCO$^+$ emission is
localized toward the OMC-3 region.
The difference in distribution between DNC and DCO$^+$
is likely due to that between N- and C-bearing molecules \citep{2008PASJ...60..407T}.
The DNC/HN$^{13}$C column density ratio 
shows high values in OMC-2 and OMC-3, and low values in OMC-1.
We guess that the OMC-2 and OMC-3 regions still contains molecular gas close to
the onset of star formation.
For the individual starless cores in Orion, the column density ratios of 
DNC/HN$^{13}$C
and DCO$^+$/H$^{13}$CO$^+$  are found to be rather constant within each core.
It may be because all the timescales of deuterization, depletion, and dynamical evolution
are comparable.
On the other hand, the core-to-core variation in the deuterium fraction is not small.
\vspace{5mm}
% \newpage
% \pagebreak

\appendix

\section{Appendix: Supplemental Tables and Figures}

Table \ref{tbl:df-int} lists
the D/H column density ratios 
(deuterium fraction) 
toward the YSOs in the northern $\int$-shaped filament
using different averaging radii, 
$r = 10\arcsec$, $r = 30\arcsec$, and $r = 1\farcm69$. 
Table \ref{tbl:df-starless} lists
the D/H column density ratios
toward the starless cores
using different averaging radii 
$r = 10\arcsec$, $r = 30\arcsec$, and the map average (All) .

\begin{figure}
\figurenum{A1}
\includegraphics[bb=0 0 505 425, width=15cm]{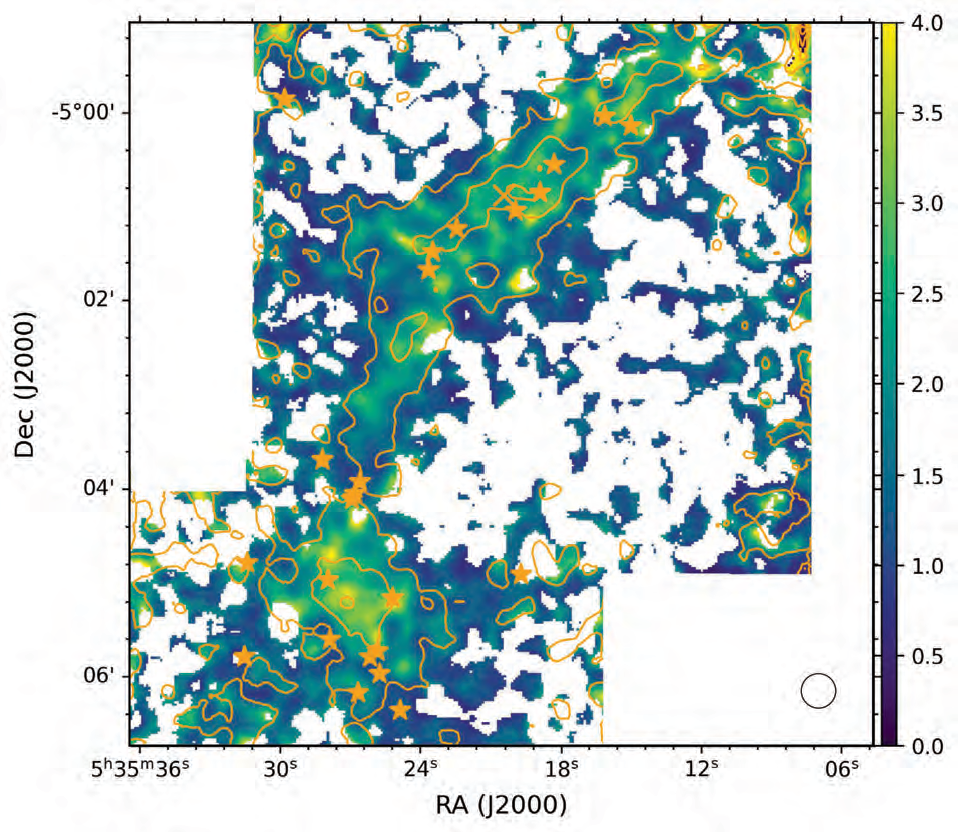}
\caption{
The color shows the column density ratio $N$(DNC)/$N$(HN$^{13}$C) toward the OMC-3 region, i.e., the upper third of the northern $\int$-shaped filament.
Contours represent the integrated-intensity distribution of DNC.
The contour interval for DNC is 1 K km s$^{-1}$ (5$\sigma$).
The orange star symbol represents \textit{Spitzer} YSOs
and HOPS sources.
The orange cross at declination $\sim -5\arcdeg 01\arcmin$ marks the position of the starless core G208.68$-$19.20N2
\citep{2024ApJ...961..123H}.
The open circle represents the HPBW for the deuterated molecules.
\label{fig:ZOADNC14N}}
\end{figure}

\begin{figure}
\figurenum{A2}
\includegraphics[bb=0 0 505 425, width=15cm]{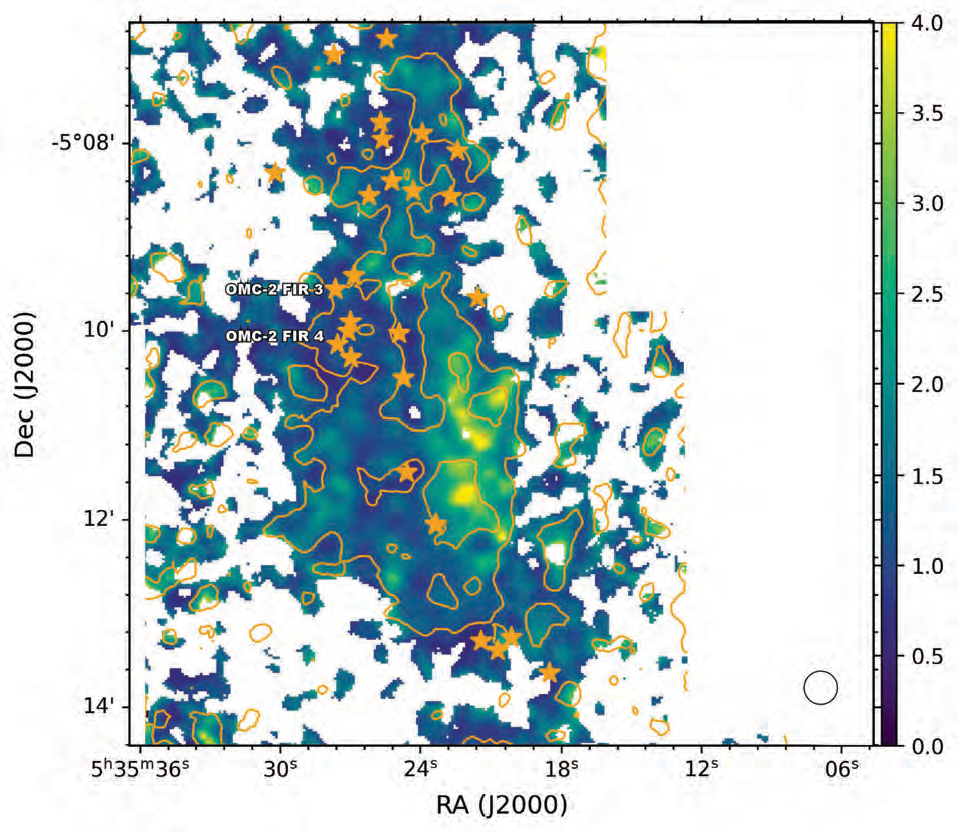}
\caption{
Same as Figure \ref{fig:ZOADNC14N}, but for the OMC-2 region, i.e., the middle third of the northern $\int$-shaped filament.
\label{fig:ZOADNC14C}}
\end{figure}

\begin{figure}
\figurenum{A3}
\includegraphics[bb=0 0 505 425, width=15cm]{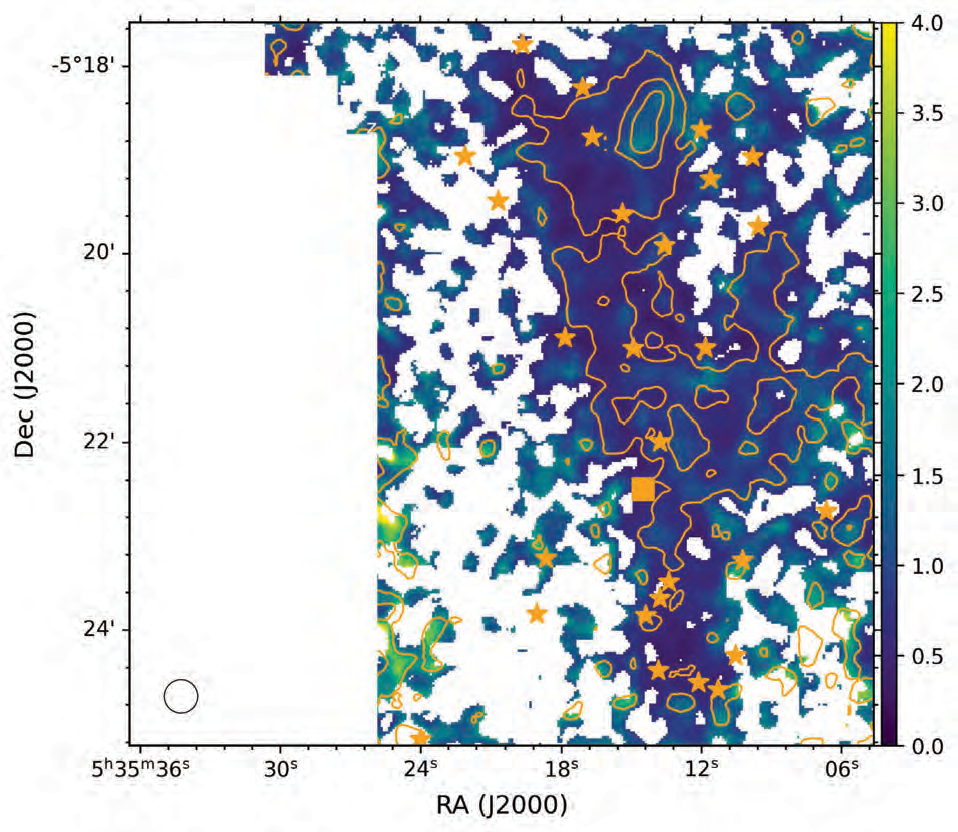}
\caption{
Same as Figure \ref{fig:ZOADNC14N}, but for the OMC-1 region, i.e., the lower third of the northern $\int$-shaped filament.
\label{fig:ZOADNC14S}}
\end{figure}

\begin{figure}
\figurenum{A4}
\includegraphics[bb=0 0 505 435, width=15cm]{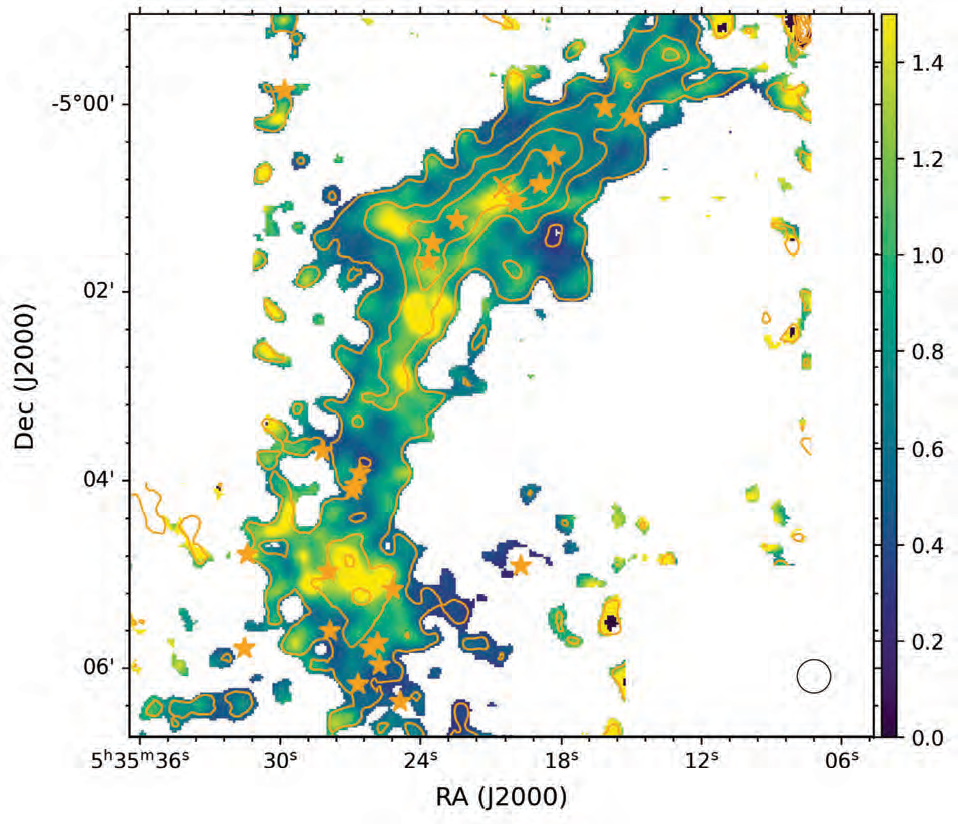}
\caption{
The color represents the column density ratio, 
$N$~(DCO$^+$)/$N$ (H$^{13}$CO$^+$)
toward the OMC-3 region, i.e., the upper third of the northern $\int$-shaped filament..
Contours represent the DCO$^+$ distribution.
The contour interval for DCO$^+$
is 1 K~km~s$^{-1}$ (5$\sigma$).
The open circle represents the HPBW for the deuterated molecules.
\label{fig:ZOADCO14N}}
\end{figure}

Figures \ref{fig:ZOADNC14N} to \ref{fig:ZOADCO14N}
illustrate the close-up views of the column density maps
toward the northern $\int$-shaped filament
separated into the three regions corresponding to
OMC-3, OMC-2, and OMC-1.
The northern $\int$-shaped filament contains many interesting sources.
Without going into details, we only highlight one starless core and one star-forming core.
G208.68-19.20N2 is a starless core located in OMC-3.
\citet{2024ApJ...961..123H} studied this region using ALMA,
and found it to be a prestellar core on the verge of the first hydrostatic core (FHSC) formation or a candidate for the FHSC.
OMC-2 FIR 3 is one of the famous YSOs in OMC-2.
It has been suggested that OMC-2 FIR 4 may have been dynamically affected by the outflow driven by OMC-2 FIR 3
\citep{2008ApJ...683..255S,2013A&A...556A..57K,2015A&A...574A.107K,2015ApJS..221...31S,2016A&A...596A..26G,2017ApJ...840...36O,2019PASJ...71S..10N,2023ApJ...944...92S}.

Figures \ref{fig:ZS04-20DNC} to \ref{fig:ZS22-32DNC}
show the column density ratio maps
of $N$~(DNC)/$N$~(HN$^{13}$C) 
toward the individual starless cores
with $T_{ex}$ = 5 K.
Figures \ref{fig:ZS04-20DCOH} to \ref{fig:ZS22-32DCOH}
show the column density ratio maps
of $N$~ (DCO$^+$)/$N$ (H$^{13}$CO$^+$) 
toward the individual starless cores
with $T_{ex}$ = 10 K.

\floattable
\begin{deluxetable}{cccccclcccccc}
\rotate
\tablecaption{Deuterium Fraction toward the YSOs in the Northern $\int$-shaped Filament \label{tbl:df-int}}
\tablecolumns{13}
\tablenum{A1}
\tablewidth{0pt}
\tablehead{
\multicolumn{3}{c}{R.A.} &
\multicolumn{3}{c}{Decl.} &
\colhead{YSO Name} &
\multicolumn{3}{c}{$N$~(DNC)/$N$~(HN$^{13}$C)} &
\multicolumn{3}{c}{$N$~(DCO$^+$)/$N$~(H$^{13}$CO$^+$)} 
\\
\cline{8-13}
\multicolumn{3}{c}{(J2000)} &
\multicolumn{3}{c}{(J2000)} &
\colhead{} &
\colhead{$r 10\arcsec$} &
\colhead{$r 30\arcsec$} &
\colhead{$r 1\farcm$69} &
\colhead{$r 10\arcsec$} &
\colhead{$r 30\arcsec$} &
\colhead{$r 1\farcm$69} 
\\
\colhead{h} &
\colhead{m} &
\colhead{s} &
\colhead{$\arcdeg$} &
\colhead{$\arcmin$} &
\colhead{$\arcsec$} &
\multicolumn{7}{c}{} 
}
\startdata
5	&	35	&	29.81 	&	-4	&	59	&	51.1 	&	HOPS 383	&	1.43 	&	1.55 	&	1.46 	&	0.74 	&	0.90 	&	0.85 	\\
5	&	35	&	16.15 	&	-5	&	0	&	2.3 	&	V* V2282 Ori	&	2.32 	&	2.27 	&	1.87 	&	0.73 	&	0.72 	&	0.77 	\\
5	&	35	&	15.03 	&	-5	&	0	&	8.2 	&	2MASS J05351503$-$0500080 	&	2.52 	&	2.09 	&	1.87 	&	0.75 	&	0.71 	&	0.77 	\\
5	&	35	&	18.32 	&	-5	&	0	&	33.0 	&	G208.68$-$19.20N3 B 	&	2.09 	&	2.17 	&	1.83 	&	0.72 	&	0.71 	&	0.78 	\\
5	&	35	&	18.91 	&	-5	&	0	&	50.9 	&	HOPS 91	&	2.52 	&	2.19 	&	1.82 	&	0.81 	&	0.82 	&	0.81 	\\
5	&	35	&	19.96 	&	-5	&	1	&	2.6 	&	2MASS J05351997$-$0501024 	&	2.71 	&	2.30 	&	1.82 	&	1.17 	&	0.91 	&	0.83 	\\
5	&	35	&	22.43 	&	-5	&	1	&	14.1 	&	HOPS 88	&	1.83 	&	2.16 	&	1.73 	&	0.97 	&	0.99 	&	0.84 	\\
5	&	35	&	23.47 	&	-5	&	1	&	28.7 	&	HOPS 87, G208.68$-$19.20North1	&	2.33 	&	2.15 	&	1.69 	&	0.94 	&	1.05 	&	0.85 	\\
5	&	35	&	23.65 	&	-5	&	1	&	40.3 	&	2MASS J05352363$-$0501402	&	2.05 	&	2.08 	&	1.69 	&	1.17 	&	1.06 	&	0.87 	\\
5	&	35	&	28.18 	&	-5	&	3	&	40.9 	&	TKK 870	&	1.59 	&	1.52 	&	1.99 	&	0.55 	&	0.73 	&	1.06 	\\
5	&	35	&	26.57 	&	-5	&	3	&	55.1 	&	G208.68$-$19.20S A	&	1.19 	&	1.69 	&	1.89 	&	0.57 	&	0.68 	&	0.94 	\\
5	&	35	&	26.81 	&	-5	&	4	&	3.0 	&	HOPS 352	&	1.57 	&	1.78 	&	1.92 	&	0.63 	&	0.71 	&	0.93 	\\
5	&	35	&	26.92 	&	-5	&	4	&	6.5 	&	2MASS J05352694$-$0504069	&	1.85 	&	1.82 	&	1.92 	&	0.62 	&	0.73 	&	0.93 	\\
5	&	35	&	31.42 	&	-5	&	4	&	47.1 	&	$[$MGM2012$]$ 2390	&	2.10 	&	1.79 	&	2.08 	&	1.15 	&	1.03 	&	1.06 	\\
5	&	35	&	19.72 	&	-5	&	4	&	54.6 	&	2MASS J05351973$-$0504544	&	1.53 	&	1.61 	&	1.68 	&	0.20 	&	0.31 	&	0.99 	\\
5	&	35	&	27.95 	&	-5	&	4	&	58.1 	&	2MASS J05352787$-$0504597	&	2.78 	&	2.62 	&	1.82 	&	1.39 	&	1.22 	&	0.85 	\\
5	&	35	&	25.18 	&	-5	&	5	&	9.4 	&	HOPS 80	&	3.04 	&	2.27 	&	1.74 	&	1.17 	&	1.00 	&	0.81 	\\
5	&	35	&	27.88 	&	-5	&	5	&	36.3 	&	2MASS J05352785$-$0505362	&	1.61 	&	2.13 	&	1.73 	&	0.56 	&	0.82 	&	0.84 	\\
5	&	35	&	25.82 	&	-5	&	5	&	43.6 	&	HOPS 78	&	2.14 	&	2.05 	&	1.69 	&	0.82 	&	0.74 	&	0.79 	\\
5	&	35	&	31.53 	&	-5	&	5	&	47.2 	&	V* V2502 Ori 	&	1.29 	&	1.45 	&	1.83 	&	0.44 	&	0.70 	&	0.95 	\\
5	&	35	&	26.16 	&	-5	&	5	&	47.3 	&	$[$KSS2017$]$ 5 	&	1.80 	&	1.97 	&	1.69 	&	0.77 	&	0.70 	&	0.79 	\\
5	&	35	&	25.75 	&	-5	&	5	&	57.9 	&	2MASS J05352574$-$0505579	&	1.46 	&	1.75 	&	1.67 	&	0.59 	&	0.64 	&	0.77 	\\
5	&	35	&	26.66 	&	-5	&	6	&	10.3 	&	HOPS 75	&	1.83 	&	1.62 	&	1.66 	&	0.55 	&	0.62 	&	0.77 	\\
5	&	35	&	24.86 	&	-5	&	6	&	21.4 	&	2MASS J05352486$-$0506216	&	1.20 	&	1.41 	&	1.59 	&	0.47 	&	0.53 	&	0.74 	\\
5	&	35	&	25.44 	&	-5	&	6	&	52.9 	&	ISOY J053525.44$-$050652.7 	&	1.10 	&	1.41 	&	1.48 	&	0.76 	&	0.59 	&	0.64 	\\
5	&	35	&	27.70 	&	-5	&	7	&	3.5 	&	HOPS 73	&	1.26 	&	1.45 	&	1.47 	&	0.76 	&	0.91 	&	0.64 	\\
5	&	35	&	25.71 	&	-5	&	7	&	46.4 	&	V* V2442 Ori	&	1.01 	&	1.18 	&	1.39 	&	…	&	0.51 	&	0.58 	\\
5	&	35	&	23.93 	&	-5	&	7	&	53.5 	&	HOPS 394	&	1.25 	&	1.30 	&	1.40 	&	0.42 	&	0.47 	&	0.60 	\\
5	&	35	&	25.61 	&	-5	&	7	&	57.3 	&	JW 774	&	0.92 	&	1.12 	&	1.39 	&	…	&	0.46 	&	0.60 	\\
5	&	35	&	22.41 	&	-5	&	8	&	4.8 	&	JW 703	&	1.37 	&	1.42 	&	1.42 	&	0.57 	&	0.48 	&	0.58 	\\
5	&	35	&	30.21 	&	-5	&	8	&	18.9 	&	2MASS J05353019$-$0508197 	&	1.42 	&	1.44 	&	1.40 	&	…	&	0.75 	&	0.70 	\\
\enddata
\end{deluxetable}

\floattable
\begin{deluxetable}{cccccclcccccc}
\rotate
\tablecaption{(continued)}
\tablecolumns{13}
\tablenum{A1}
\tablewidth{0pt}
\tablehead{
\multicolumn{3}{c}{R.A.} &
\multicolumn{3}{c}{Decl.} &
\colhead{YSO Name} &
\multicolumn{3}{c}{$N$~(DNC)/$N$~(HN$^{13}$C)} &
\multicolumn{3}{c}{$N$~(DCO$^+$)/$N$~(H$^{13}$CO$^+$)} 
\\
\cline{8-13}
\multicolumn{3}{c}{(J2000)} &
\multicolumn{3}{c}{(J2000)} &
\colhead{} &
\colhead{$r 10\arcsec$} &
\colhead{$r 30\arcsec$} &
\colhead{$r 1\farcm$69} &
\colhead{$r 10\arcsec$} &
\colhead{$r 30\arcsec$} &
\colhead{$r 1\farcm$69} 
\\
\colhead{h} &
\colhead{m} &
\colhead{s} &
\colhead{$\arcdeg$} &
\colhead{$\arcmin$} &
\colhead{$\arcsec$} &
\multicolumn{7}{c}{} 
}
\startdata
5	&	35	&	25.22 	&	-5	&	8	&	24.0 	&	HOPS 69	&	1.38 	&	1.38 	&	1.39 	&	0.49 	&	0.56 	&	0.58 	\\
5	&	35	&	24.30 	&	-5	&	8	&	30.6 	&	HOPS 68	&	1.78 	&	1.40 	&	1.40 	&	0.68 	&	0.59 	&	0.58 	\\
5	&	35	&	26.20 	&	-5	&	8	&	33.4 	&	HOPS 349	&	1.45 	&	1.44 	&	1.38 	&	0.40 	&	0.55 	&	0.59 	\\
5	&	35	&	22.69 	&	-5	&	8	&	34.0 	&	HOPS 67	&	1.27 	&	1.40 	&	1.43 	&	0.54 	&	0.61 	&	0.62 	\\
5	&	35	&	26.84 	&	-5	&	9	&	24.6 	&	V* V2455 Ori 	&	1.34 	&	1.35 	&	1.35 	&	0.52 	&	0.52 	&	0.67 	\\
5	&	35	&	27.63 	&	-5	&	9	&	33.5 	&	2MASS J05352762$-$0509337, OMC-2 FIR 3	&	1.19 	&	1.23 	&	1.36 	&	…	&	0.52 	&	0.72 	\\
5	&	35	&	21.55 	&	-5	&	9	&	38.7 	&	V* V2377 Ori	&	1.29 	&	1.52 	&	1.58 	&	…	&	0.71 	&	0.84 	\\
5	&	35	&	27.00 	&	-5	&	9	&	54.1 	&	V* V2457 Ori, OMC-2 FIR 4c	&	1.06 	&	1.08 	&	1.42 	&	0.42 	&	0.51 	&	0.77 	\\
5	&	35	&	27.07 	&	-5	&	10	&	0.4 	&	HOPS 108, OMC-2 FIR 4h	&	1.07 	&	1.01 	&	1.42 	&	0.35 	&	0.51 	&	0.81 	\\
5	&	35	&	24.90 	&	-5	&	10	&	1.5 	&	2MASS J05352485$-$0510016	&	0.86 	&	1.20 	&	1.51 	&	…	&	0.85 	&	0.82 	\\
5	&	35	&	27.56 	&	-5	&	10	&	8.7 	&	2MASS J05352755$-$0510083, OMC-2 FIR 4j	&	1.00 	&	0.97 	&	1.41 	&	0.43 	&	0.55 	&	0.89 	\\
5	&	35	&	26.97 	&	-5	&	10	&	17.2 	&	2MASS J05352696$-$0510173, OMC-2 FIR 4k	&	1.02 	&	1.00 	&	1.46 	&	0.38 	&	0.54 	&	0.88 	\\
5	&	35	&	24.73 	&	-5	&	10	&	30.2 	&	2MASS J05352477$-$0510296	&	1.33 	&	1.28 	&	1.55 	&	…	&	0.75 	&	0.89 	\\
5	&	35	&	24.58 	&	-5	&	11	&	29.7 	&	V* V2427 Ori	&	0.94 	&	1.40 	&	1.61 	&	…	&	1.17 	&	0.95 	\\
5	&	35	&	23.33 	&	-5	&	12	&	3.1 	&	HOY J053523.29$-$051203.0 	&	1.58 	&	1.76 	&	1.60 	&	…	&	0.87 	&	0.95 	\\
5	&	35	&	20.14 	&	-5	&	13	&	15.5 	&	V* V2364 Ori 	&	1.65 	&	1.32 	&	1.51 	&	0.94 	&	0.80 	&	0.95 	\\
5	&	35	&	21.40 	&	-5	&	13	&	17.5 	&	HOPS 409	&	0.85 	&	1.16 	&	1.49 	&	…	&	0.90 	&	0.95 	\\
5	&	35	&	20.73 	&	-5	&	13	&	23.6 	&	2MASS J05352076$-$0513225	&	1.37 	&	1.25 	&	1.47 	&	0.99 	&	0.84 	&	0.93 	\\
5	&	35	&	18.51 	&	-5	&	13	&	38.2 	&	V* V2331 Ori	&	1.14 	&	1.46 	&	1.39 	&	0.86 	&	0.90 	&	0.94 	\\
5	&	35	&	19.84 	&	-5	&	15	&	8.5 	&	V* V2359 Ori 	&	0.94 	&	1.12 	&	1.21 	&	…	&	0.84 	&	0.94 	\\
5	&	35	&	19.47 	&	-5	&	15	&	32.7 	&	HOPS 56	&	0.98 	&	1.14 	&	1.20 	&	0.51 	&	0.80 	&	0.94 	\\
5	&	35	&	19.66 	&	-5	&	17	&	46.2 	&	$[$MGM2012$]$ 2106	&	0.53 	&	0.69 	&	0.94 	&	0.55 	&	0.46 	&	0.37 	\\
5	&	35	&	17.09 	&	-5	&	18	&	13.9 	&	2MASS J05351712$-$0518137	&	0.78 	&	0.71 	&	0.86 	&	0.37 	&	0.26 	&	0.42 	\\
5	&	35	&	12.02 	&	-5	&	18	&	40.8 	&	V* V2222 Ori	&	1.29 	&	1.16 	&	0.91 	&	…	&	0.33 	&	0.45 	\\
5	&	35	&	16.69 	&	-5	&	18	&	45.2 	&	2MASS J05351671$-$0518448	&	0.69 	&	0.70 	&	0.82 	&	0.10 	&	0.13 	&	0.37 	\\
5	&	35	&	22.10 	&	-5	&	18	&	57.7 	&	2MASS J05352210$-$0518577	&	0.90 	&	0.99 	&	0.93 	&	…	&	…	&	0.75 	\\
5	&	35	&	9.82 	&	-5	&	18	&	58.1 	&	V* V2192 Ori	&	0.71 	&	0.90 	&	0.97 	&	…	&	…	&	0.59 	\\
5	&	35	&	11.61 	&	-5	&	19	&	12.4 	&	2MASS J05351162$-$0519122	&	1.11 	&	0.87 	&	0.86 	&	…	&	…	&	0.45 	\\
5	&	35	&	20.71 	&	-5	&	19	&	26.3 	&	$[$AD95$]$ 1362	&	0.63 	&	0.91 	&	0.87 	&	…	&	0.80 	&	0.73 	\\
5	&	35	&	15.39 	&	-5	&	19	&	34.4 	&	COUP 702 	&	0.53 	&	0.57 	&	0.76 	&	…	&	0.15 	&	0.19 	\\
5	&	35	&	9.57 	&	-5	&	19	&	42.7 	&	2MASS J05350957$-$0519426	&	0.71 	&	0.83 	&	0.89 	&	…	&	…	&	0.58 	\\
5	&	35	&	13.60 	&	-5	&	19	&	54.9 	&	2MASS J05351362$-$0519548	&	0.90 	&	0.69 	&	0.78 	&	…	&	…	&	0.17 	\\
5	&	35	&	17.84 	&	-5	&	20	&	53.9 	&	V* V2314 Ori A 	&	0.52 	&	0.71 	&	0.80 	&	…	&	0.12 	&	0.39 	\\
\enddata
\end{deluxetable}

\floattable
\begin{deluxetable}{cccccclcccccc}
\rotate
\tablecaption{(continued)}
\tablecolumns{13}
\tablenum{A1}
\tablewidth{0pt}
\tablehead{
\multicolumn{3}{c}{R.A.} &
\multicolumn{3}{c}{Decl.} &
\colhead{YSO Name} &
\multicolumn{3}{c}{$N$~(DNC)/$N$~(HN$^{13}$C)} &
\multicolumn{3}{c}{$N$~(DCO$^+$)/$N$~(H$^{13}$CO$^+$)} 
\\
\cline{8-13}
\multicolumn{3}{c}{(J2000)} &
\multicolumn{3}{c}{(J2000)} &
\colhead{} &
\colhead{$r 10\arcsec$} &
\colhead{$r 30\arcsec$} &
\colhead{$r 1\farcm$69} &
\colhead{$r 10\arcsec$} &
\colhead{$r 30\arcsec$} &
\colhead{$r 1\farcm$69}
\\
\colhead{h} &
\colhead{m} &
\colhead{s} &
\colhead{$\arcdeg$} &
\colhead{$\arcmin$} &
\colhead{$\arcsec$} &
\multicolumn{7}{c}{} 
}
\startdata
5	&	35	&	11.84 	&	-5	&	21	&	0.3 	&	COUP J053511.8-052100	&	0.64 	&	0.77 	&	0.80 	&	…	&	0.18 	&	0.24 	\\
5	&	35	&	14.93 	&	-5	&	21	&	0.6 	&	2MASS J05351495-0521009 	&	0.61 	&	0.63 	&	0.77 	&	…	&	0.15 	&	0.34 	\\
5	&	35	&	13.79 	&	-5	&	21	&	59.9 	&	V* V2248 Ori	&	0.81 	&	0.75 	&	0.87 	&	…	&	0.21 	&	0.53 	\\
5	&	35	&	4.15 	&	-5	&	22	&	32.0 	&	V* V2137 Ori	&	0.63 	&	0.84 	&	1.16 	&	0.48 	&	0.72 	&	1.10 	\\
5	&	35	&	6.66 	&	-5	&	22	&	44.2 	&	MLLA 556	&	0.91 	&	1.06 	&	1.09 	&	…	&	0.54 	&	0.99 	\\
5	&	35	&	18.66 	&	-5	&	23	&	14.1 	&	V* AF Ori	&	1.72 	&	1.46 	&	1.16 	&	1.51 	&	1.57 	&	0.96 	\\
5	&	35	&	10.25 	&	-5	&	23	&	15.5 	&	2MASS J05351026-0523163	&	1.51 	&	1.17 	&	1.02 	&	0.58 	&	0.59 	&	0.92 	\\
5	&	35	&	13.40 	&	-5	&	23	&	29.2 	&	ISOY J053513.41-052329.3	&	0.57 	&	0.76 	&	1.04 	&	…	&	0.26 	&	0.66 	\\
5	&	35	&	13.78 	&	-5	&	23	&	40.0 	&	ISOY J053513.81-052340.1	&	0.60 	&	0.75 	&	1.06 	&	…	&	0.25 	&	0.71 	\\
5	&	35	&	14.38 	&	-5	&	23	&	50.9 	&	$[$ZRK2004b$]$ 144-351 	&	0.85 	&	0.73 	&	1.08 	&	…	&	0.24 	&	0.79 	\\
5	&	35	&	10.54 	&	-5	&	24	&	16.5 	&	V* V2202 Ori	&	0.95 	&	1.13 	&	1.16 	&	…	&	0.49 	&	1.13 	\\
5	&	35	&	13.84 	&	-5	&	24	&	26.1 	&	COUP 583 	&	0.51 	&	0.71 	&	1.10 	&	…	&	0.39 	&	1.09 	\\
5	&	35	&	12.12 	&	-5	&	24	&	33.8 	&	V* V2224 Ori 	&	0.58 	&	0.84 	&	1.14 	&	…	&	0.30 	&	1.03 	\\
5	&	35	&	11.30 	&	-5	&	24	&	38.3 	&	2MASS J05351131-0524381	&	0.88 	&	0.99 	&	1.19 	&	0.35 	&	0.72 	&	1.15 	\\
5	&	35	&	23.98 	&	-5	&	25	&	9.9 	&	2MASS J05352398-0525098	&	1.74 	&	1.57 	&	1.95 	&	1.44 	&	1.48 	&	1.94 	\\
5	&	35	&	19.37 	&	-5	&	25	&	42.1 	&	V* V2350 Ori	&	1.19 	&	1.49 	&	1.36 	&	1.44 	&	1.58 	&	1.47 	\\
\hline																									
mean	&		&		&		&		&		&		&	1.31 	&	1.35 	&	1.39 	&	0.71 	&	0.67 	&	0.78 	\\
stdev	&		&		&		&		&		&		&	0.58 	&	0.49 	&	0.35 	&	0.33 	&	0.31 	&	0.27 	\\
\enddata
\end{deluxetable}

\newpage

\floattable
\begin{deluxetable}{cccccccccccccccccc}
\rotate
\tablecaption{Deuterium Fraction toward the starless cores \label{tbl:df-starless}}
\tablecolumns{18}
\tablenum{A2}
\tablewidth{0pt}
\tablehead{
\colhead{Core \#} &
\colhead{Core Name} &
\multicolumn{3}{c}{R.A.} &
\multicolumn{3}{c}{Decl.} &
\multicolumn{6}{c}{$N$~(DNC)/$N$~(HN$^{13}$C)} &
\multicolumn{3}{c}{$N$~(DCO$^+$)/$N$~(H$^{13}$CO$^+$)} &
\colhead{$N$~(DNC)/$N$~(HN$^{13}$C)}
\\
\cline{9-17}
\colhead{} &
\colhead{} &
\multicolumn{3}{c}{(J2000)} &
\multicolumn{3}{c}{(J2000)} &
\multicolumn{3}{c}{$T_{ex}$ = 5 K} &
\multicolumn{3}{c}{$T_{ex}$ = 10 K} &
\multicolumn{3}{c}{$T_{ex}$ = 10 K} &
\colhead{\citet{2020ApJS..249...33K}}
\\
\cline{9-17}
\colhead{} &
\colhead{} &
\colhead{} &
\colhead{} &
\colhead{} &
\colhead{} &
\colhead{} &
\colhead{} &
\colhead{$r 10\arcsec$} &
\colhead{$r 30\arcsec$} &
\colhead{All} &
\colhead{$r 10\arcsec$} &
\colhead{$r 30\arcsec$} &
\colhead{All} &
\colhead{$r 10\arcsec$} &
\colhead{$r 30\arcsec$} &
\colhead{All} &
\colhead{}
\\
\colhead{} &
\colhead{} &
\colhead{h} &
\colhead{m} &
\colhead{s} &
\colhead{$\arcdeg$} &
\colhead{$\arcmin$} &
\colhead{$\arcsec$} &
\colhead{} &
\colhead{} &
\colhead{} &
\colhead{} &
\colhead{} &
\colhead{} &
\colhead{} &
\colhead{} &
\colhead{} &
\colhead{} 
}
\startdata
04	&	G201.72$-$11.22	&	5	&	50	&	54.53 	&	4	&	37	&	42.6 	&	3.66 	&	3.79 	&	3.05 	&	3.74 	&	3.79 	&	3.11 	&	4.60 	&	4.70 	&	3.48 	&		\\
10	&	G206.93$-$16.61East1	&	5	&	41	&	40.54 	&	-2	&	17	&	4.3 	&	2.96 	&	2.05 	&	1.14 	&	2.61 	&	1.86 	&	1.13 	&	1.12 	&	1.33 	&	1.64 	&		\\
11	&	G206.93$-$16.61West4	&	5	&	41	&	25.84 	&	-2	&	19	&	28.4 	&	2.50 	&	2.22 	&	1.45 	&	2.46 	&	2.19 	&	1.45 	&	2.85 	&	2.58 	&	1.96 	&	5.1 	\\
12	&	G206.93$-$16.61West5	&	5	&	41	&	28.77 	&	-2	&	20	&	4.3 	&	1.82 	&	1.74 	&	1.45 	&	1.86 	&	1.79 	&	1.45 	&	3.54 	&	2.64 	&	1.96 	&		\\
14	&	G207.36$-$19.82North4	&	5	&	30	&	44.81 	&	-4	&	10	&	27.6 	&	2.68 	&	2.03 	&	1.40 	&	2.27 	&	1.86 	&	1.33 	&	8.73 	&	5.95 	&	4.15 	&		\\
15	&	G207.36$-$19.82South	&	5	&	30	&	46.81 	&	-4	&	12	&	29.4 	&	1.98 	&	2.50 	&	1.39 	&	1.94 	&	2.38 	&	1.40 	&	6.36 	&	5.56 	&	3.64 	&		\\
17	&	G209.05$-$19.73North	&	5	&	34	&	3.96 	&	-5	&	32	&	42.5 	&	6.10 	&	2.98 	&	2.61 	&	5.00 	&	2.75 	&	2.48 	&	2.13 	&	2.51 	&	2.83 	&	5.7 	\\
18	&	G209.05$-$19.73South	&	5	&	34	&	3.12 	&	-5	&	34	&	11.0 	&	8.41 	&	6.59 	&	4.62 	&	6.95 	&	5.72 	&	4.19 	&	6.61 	&	6.31 	&	3.58 	&	5.9 	\\
19	&	G209.29$-$19.65North1	&	5	&	35	&	0.25 	&	-5	&	40	&	2.4 	&	4.58 	&	3.28 	&	2.66 	&	5.06 	&	3.56 	&	2.71 	&	1.61 	&	1.73 	&	1.68 	&	2.0 	\\
20	&	G209.29$-$19.65South1	&	5	&	34	&	55.99 	&	-5	&	46	&	3.2 	&	2.47 	&	1.72 	&	1.05 	&	2.32 	&	1.69 	&	1.08 	&	2.62 	&	2.21 	&	2.19 	&		\\
21	&	G209.29$-$19.65South2	&	5	&	34	&	53.81 	&	-5	&	46	&	12.8 	&	2.13 	&	1.43 	&	1.05 	&	2.05 	&	1.43 	&	1.08 	&	2.17 	&	2.11 	&	2.19 	&		\\
22	&	G209.55$-$19.68North2	&	5	&	35	&	7.01 	&	-5	&	56	&	38.4 	&	1.15 	&	1.30 	&	1.67 	&	1.88 	&	1.54 	&	1.74 	&	6.62 	&	6.93 	&	7.45 	&		\\
23	&	G209.77$-$19.40West	&	5	&	36	&	21.19 	&	-6	&	1	&	32.7 	&	2.89 	&	1.92 	&	1.67 	&	2.74 	&	1.94 	&	1.72 	&	5.93 	&	4.93 	&	3.87 	&		\\
24	&	G209.77$-$19.40East2	&	5	&	36	&	32.19 	&	-6	&	2	&	4.7 	&	9.09 	&	9.95 	&	8.24 	&	8.73 	&	8.76 	&	7.42 	&	6.55 	&	5.22 	&	3.91 	&	2.5 	\\
25	&	G209.77$-$19.40East3	&	5	&	36	&	35.94 	&	-6	&	2	&	44.7 	&	12.40 	&	9.45 	&	8.24 	&	9.45 	&	8.26 	&	7.42 	&	6.91 	&	5.81 	&	3.91 	&	4.7,1.9	\\
27	&	G209.79$-$19.80West	&	5	&	35	&	11.19 	&	-6	&	14	&	0.7 	&	6.25 	&	4.51 	&	2.98 	&	4.74 	&	4.00 	&	2.80 	&	4.83 	&	3.59 	&	2.78 	&		\\
28	&	G209.94$-$19.52South1	&	5	&	36	&	24.96 	&	-6	&	14	&	4.7 	&	6.96 	&	9.73 	&	5.53 	&	6.33 	&	8.00 	&	5.07 	&	4.88 	&	4.55 	&	3.60 	&	6.3,4.1	\\
29	&	G210.37$-$19.53North	&	5	&	36	&	55.03 	&	-6	&	34	&	33.2 	&	. . .	&	. . .	&	. . .	&	. . .	&	. . .	&	1.03 	&	5.18 	&	5.05 	&	4.23 	&		\\
30	&	G210.82$-$19.47North2	&	5	&	37	&	59.84 	&	-6	&	57	&	9.9 	&	1.79 	&	2.15 	&	1.60 	&	1.68 	&	1.97 	&	1.52 	&	2.91 	&	3.46 	&	3.49 	&	5.3 	\\
32	&	G211.16$-$19.33North3	&	5	&	39	&	2.26 	&	-7	&	11	&	7.9 	&	13.19 	&	7.93 	&	5.78 	&	9.74 	&	7.41 	&	5.67 	&	4.87 	&	3.86 	&	2.75 	&	$<$3.5	\\
\enddata
\tablecomments{Two numerals in \citet{2020ApJS..249...33K} represent 
values for multiple velocity components.}
\end{deluxetable}

\newpage

\newpage

\newpage

\newpage

\begin{figure}
\figurenum{A5}
\includegraphics[bb=0 0 505 825, width=10cm]{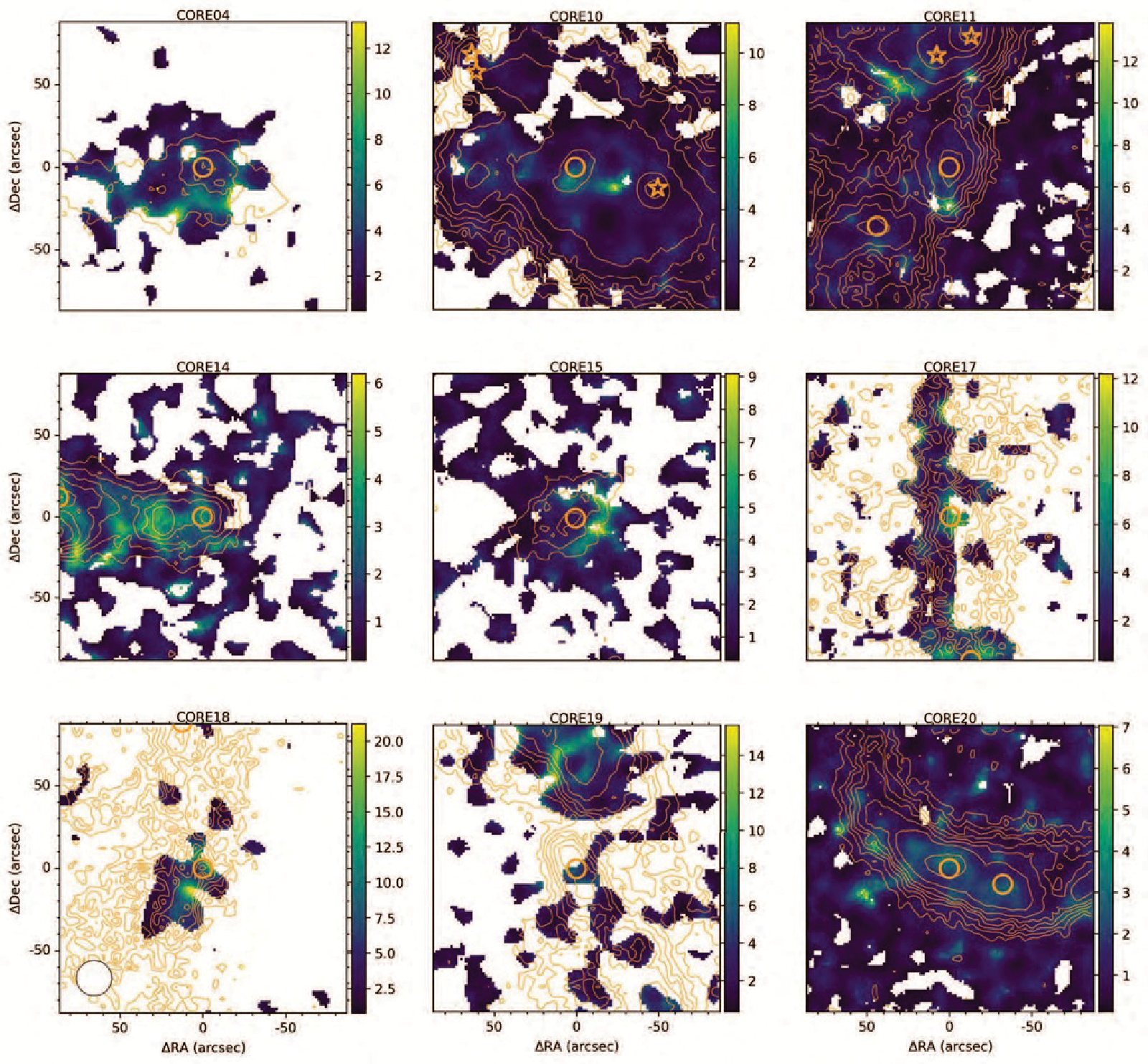}
\caption{The color shows the column density ratio $N$(DNC)/$N$(HN$^{13}$C) toward 
the individual starless cores, calculated assuming $T_{\rm ex}$ = 5 K.
Contours represent the same 850 $\mu$m continuum distribution as in Figure \ref{fig:ZS04-20DNC_10}.
The open circle represents the HPBW for the deuterated molecules.
\label{fig:ZS04-20DNC}}
\end{figure}
\newpage

\begin{figure}
\figurenum{A5}
\includegraphics[bb=0 0 505 825, width=10cm]{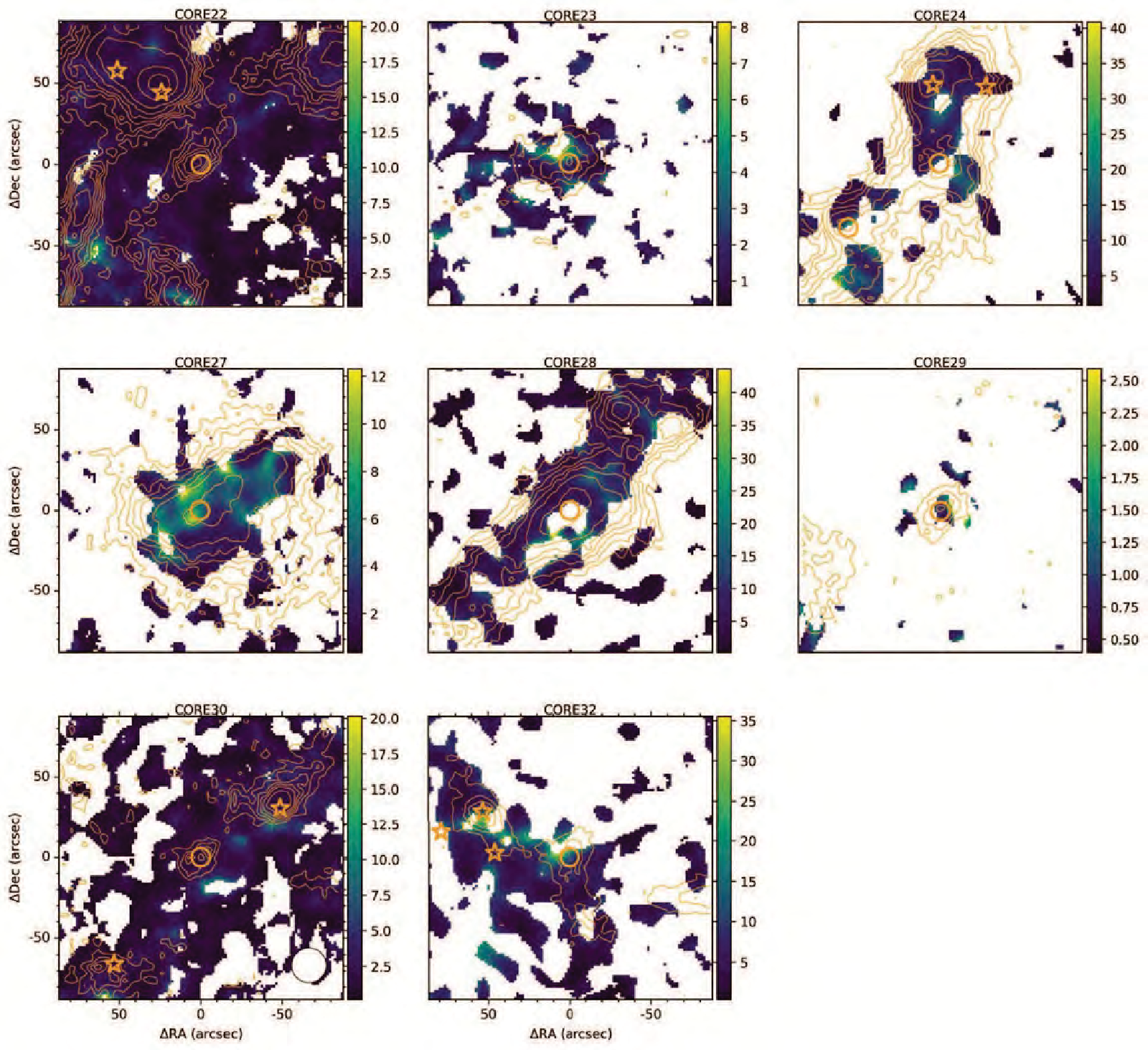}
\caption{Continued. \label{fig:ZS22-32DNC}}
\end{figure}
\newpage

\begin{figure}
\figurenum{A6}
\includegraphics[bb=0 0 505 825, width=10cm]{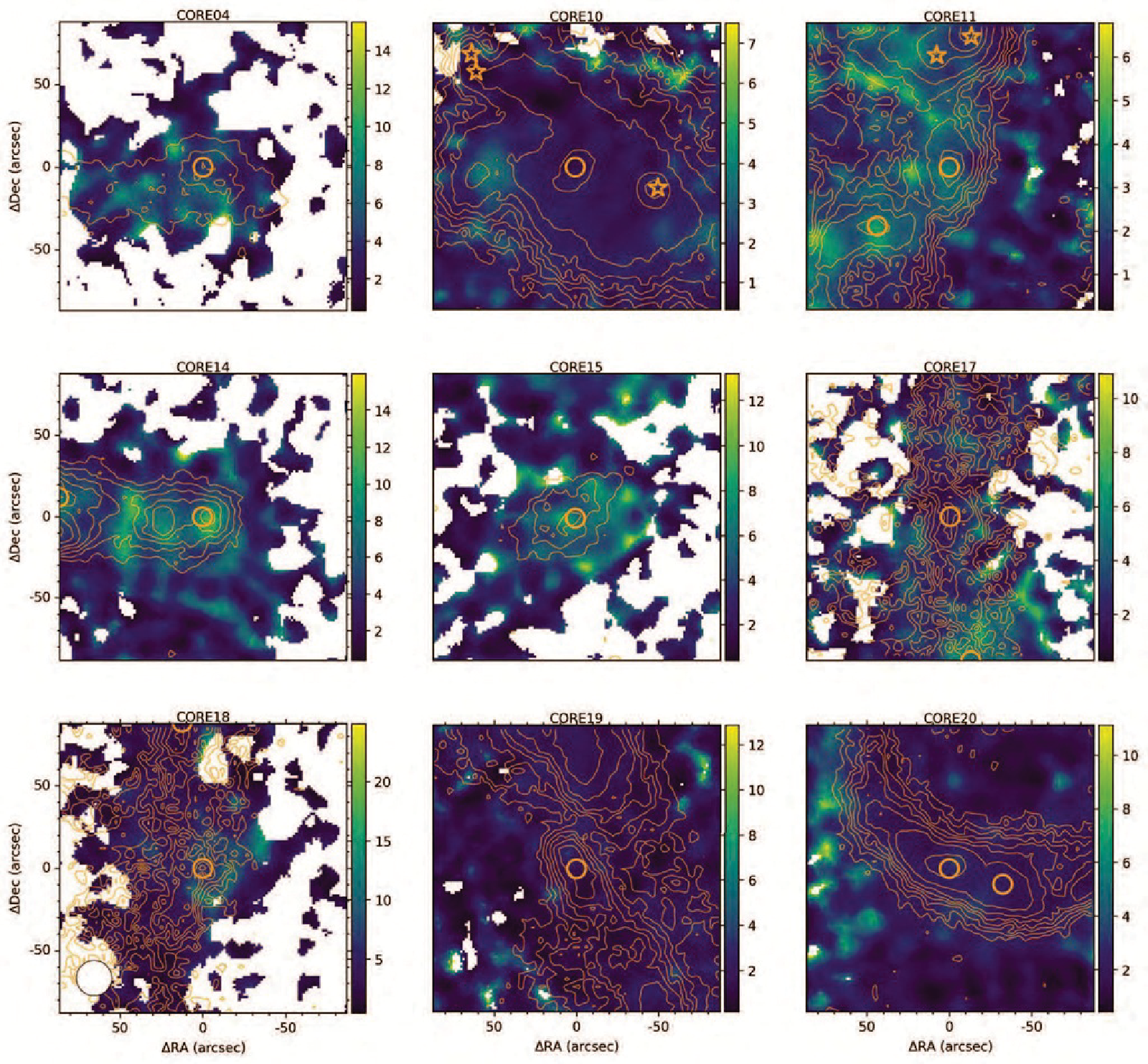}
\caption{
The color shows the column density ratio $N$~(DCO$^+$)/$N$ (H$^{13}$CO$^+$) toward 
the individual starless cores, calculated assuming $T_{\rm ex}$ = 10 K.
Contours represent the same 850 $\mu$m continuum distribution as in Figure \ref{fig:ZS04-20DNC_10}.
The open circle represents the HPBW for the deuterated molecules.
\label{fig:ZS04-20DCOH}}
\end{figure}
\newpage

\begin{figure}
\figurenum{A6}
\includegraphics[bb=0 0 505 825, width=10cm]{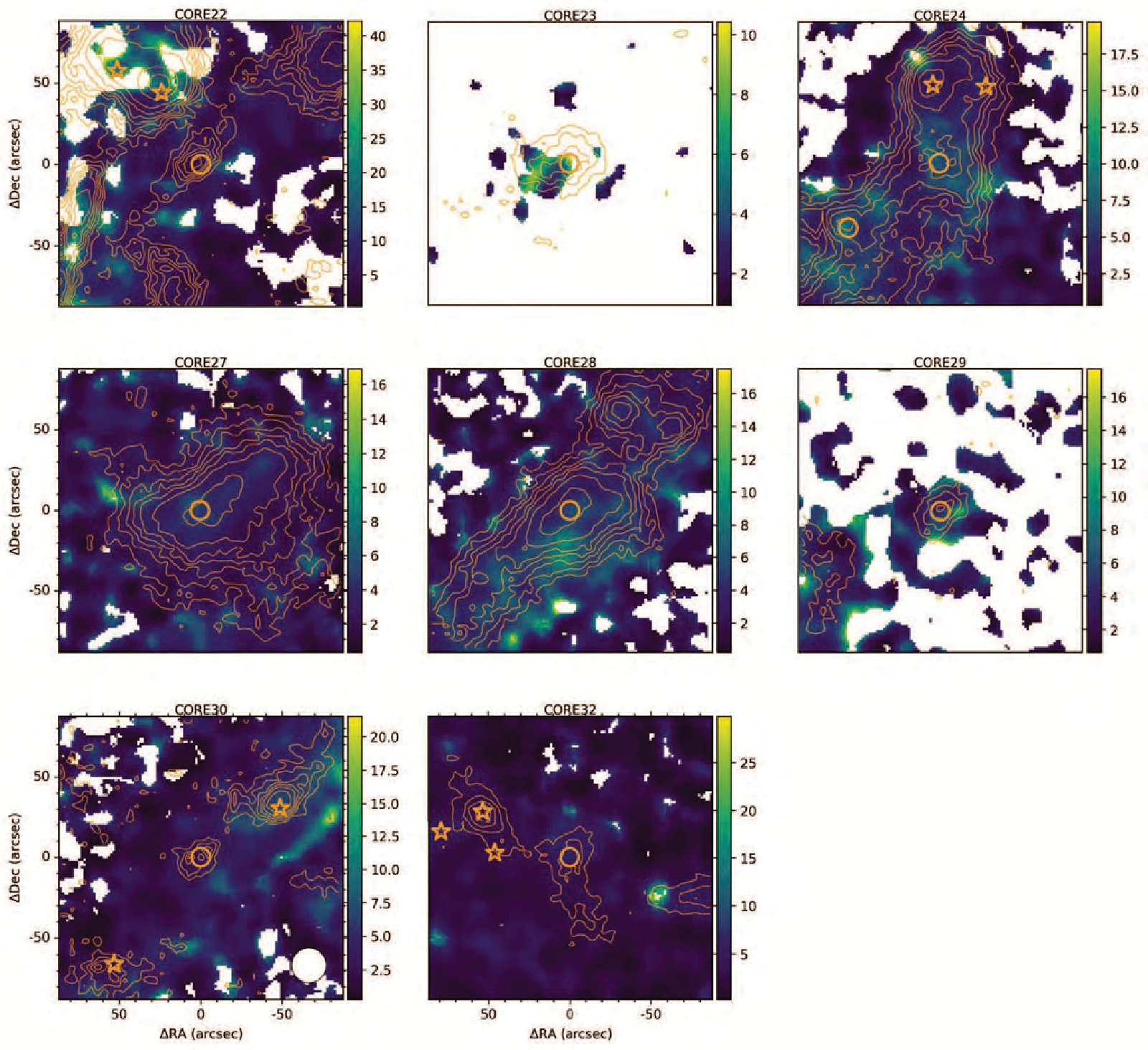}
\caption{Continued. \label{fig:ZS22-32DCOH}}
\end{figure}
\newpage

%\bibliography{ref}  

\begin{acknowledgments}
Data analysis was carried out using the Multi-wavelength Data Analysis System operated by the Astronomy Data Center (ADC), 
National Astronomical Observatory of Japan.
This work was supported by JSPS KAKENHI (Grant Number JP20H05645). 
D.S. acknowledges the support from Ramanujan Fellowship (ANRF,RJF/2021/000116) and PRL, India.
S.-Y. H acknowledges supports from the Academia Sinica of Taiwan (grant No. AS-PD-1142-M02-2) 
and National Science and Technology Council of Taiwan (grant No. 112-2112-M-001- 039-MY3).  
\end{acknowledgments}

\vspace{5mm}
\facilities{No:45m}

\software{AIPS \citep{1996ASPC..101...37V}, astropy (Astropy Collaboration et al. 2013, 2018), NOSTAR \citep{2008PASJ...60..445S}}

\end{document}